\documentclass[prb,showpacs,twocolumn,aps,superscriptaddress,a4paper]{revtex4-1}

\usepackage{color}
\usepackage{dcolumn}
\usepackage{amsmath}
\usepackage{amssymb}
\usepackage{mathtools}
\usepackage{graphicx}
\usepackage{latexsym}
\usepackage{amsfonts}
\usepackage{amssymb}
\DeclareGraphicsExtensions{.pdf,.gif,.jpg}

\newcommand{\be}{\begin{equation}}  
\newcommand{\ee}{\end{equation}}
\newcommand{\beq}{\begin{eqnarray}}  
\newcommand{\eeq}{\end{eqnarray}}  
\renewcommand\Im{\operatorname{Im}}
\renewcommand\Re{\operatorname{Re}}
\def\ud{{\mathrm{d}}}
\def\eps{{\epsilon}}
\def\g{\gamma}
\def\w{{\omega}}
\def\im{{\mathrm{i}}}
\def\ex{{\mathrm{e}}}
\def\intw{{\int_{-\infty}^\infty\hspace{-2pt}\frac{\ud \omega}{2\pi}}}
\def\balpha{\mbox{\boldmath $\alpha$}}
\def\brho{\mbox{\boldmath $\rho$}}
\def\unity{\mbox{\boldmath $1$}}
\def\bim{\mbox{$\im\unity$}}
\def\bd{\mbox{\boldmath $d$}}
\def\bD{\mbox{\boldmath $D$}}
\def\bB{\mbox{\boldmath $B$}}
\def\bw{\mbox{\boldmath $\w$}}
\def\bphi{\mbox{\boldmath $\phi$}}
\def\bOmega{\mbox{\boldmath $\varOmega$}}
\def\bPi{\mbox{\boldmath $\varPi$}}
\def\bLambda{\mbox{\boldmath $\varLambda$}}
\def\bGamma{\mbox{\boldmath $\varGamma$}}
\def\bI{\mbox{\boldmath $I$}}
\def\tb{\bar{t}}

\begin{document}


\title{Phononic heat transport in the transient regime: An analytic solution}

\author{Riku Tuovinen}
\affiliation{Department of Physics, Nanoscience Center, FIN 40014, 
             University of Jyv{\"a}skyl{\"a}, Finland}

\author{Niko S{\"a}kkinen}
\affiliation{Department of Physics, Nanoscience Center, FIN 40014, 
             University of Jyv{\"a}skyl{\"a}, Finland}

\author{Daniel Karlsson}
\affiliation{Department of Physics, Nanoscience Center, FIN 40014, 
             University of Jyv{\"a}skyl{\"a}, Finland}

\author{Gianluca Stefanucci}
\affiliation{Dipartimento di Fisica, Universit\`a di Roma Tor Vergata, 
             Via della Ricerca Scientifica 1, I-00133 Rome, Italy, and European Theoretical Spectroscopy Facility (ETSF)}
\affiliation{Laboratori Nazionali di Frascati, Istituto Nazionale di Fisica Nucleare, 
             Via E. Fermi 40, 00044 Frascati, Italy}

\author{Robert van Leeuwen}
\affiliation{Department of Physics, Nanoscience Center, FIN 40014, 
             University of Jyv{\"a}skyl{\"a}, Finland, and European Theoretical Spectroscopy Facility (ETSF)}

\date{\today}  

\begin{abstract}
We investigate the time-resolved quantum transport properties of phonons in arbitrary harmonic systems connected to phonon baths at different temperatures. We obtain a closed analytic expression of the time-dependent one-particle reduced density matrix by explicitly solving the equations of motion for the nonequilibrium Green's function. This is achieved through a well-controlled approximation of the frequency-dependent bath self-energy. Our result allows for exploring transient oscillations and relaxation times of local heat currents, and correctly reduces to an earlier known result in the steady-state limit. We apply the formalism to atomic chains, and benchmark the validity of the approximation against full numerical solutions of the bosonic Kadanoff--Baym equations for the Green's function. We find good agreement between the analytic and numerical solutions for weak contacts and baths with a wide energy dispersion. We further analyze relaxation times from low to high temperature gradients.
\end{abstract}

  
\maketitle  


\section{Introduction}\label{sec:intro}

Heat in a nanoscale junction is transported by electrons and molecular vibrations known as \emph{phonons} or \emph{vibrons}. In traversing an electronic device, an electron may transfer a portion of its energy to the structure's molecular vibrations thus creating a heat wavefront propagating through the system\cite{Segal2003,Galperin2007,RevModPhys.83.131}. Therefore, phonons are ubiquitous in virtually any molecular or nanoscale junction. In spite of the progress and advances in studying electron transport phenomena in nanoscale structures since the early 1970s\cite{Aviram1974}, measurements of thermal conductance at this scale presents several technical and conceptual difficulties and became available only recently. Suspended nanostructures such as monochrystalline GaAs\cite{Tighe1997} and silicon nitride layers\cite{Schwab2000} have been among the first settings where quantized thermal conductance was observed. Further thermal conductance studies have been performed in carbon nanotubes\cite{Hone1999, Yang2002} and silicon nanowires\cite{Deyu2003, Bourgeois2007}, and ultimately reaching the scale of molecular and atomic contacts\cite{Ge2006, Wang2007exp, Feher2007}.

A considerable amount of purely electronic transport simulations has been done according to the intuitive framework by Landauer\cite{Landauer1957} and B{\"u}ttiker\cite{Buttiker1986}. The basic idea is that the charge dynamics in a conducting channel connected to electrodes is governed by a scattering process where charge carriers are either transmitted between the electrodes through the channel, or reflected back to an electrode from the channel. From this approach an expression for the current through the channel is obtained in terms of transmission probabilities. This framework has been particularly useful after the more rigorous work by \citet{Caroli1971-1,Caroli1971-2} based on the nonequilibrium Green's functions. Further generalizations to the formalism are due to \citet{MeirWingreen1992,Pastawski1992,Jauho1994,Stefanucci2004,PhysRevB.80.115107} who discussed electron interactions in the conducting channel, dissipation due to many-body effects, time-dependent voltages and fields, and the role of initial correlations. In the case of noninteracting particles the same methodology has been proposed up to the derivation of a time-dependent Landauer--B{\"u}ttiker formula in simple systems\cite{Stefanucci2004,Perfetto2008} and arbitrary structures with time-dependent bias voltages\cite{Tuovinen2013,Tuovinen2014,Ridley2015}.

The Landauer--B{\"u}ttiker formalism is not limited to electronic transport. It also applies in the context of thermal transport for phononic systems as proposed by \citet{Rego1998,Rego2001,Segal2003} via the wave scattering method (similar to the original work of Landauer) and by \citet{PhysRevLett.96.255503,Wang2006,Wang2007} in terms of nonequilibrium Green's functions (similar to the work of Meir and Wingreen). In both approaches, expressions for the thermal conductance and the heat current are recovered as a Landauer--B{\"u}ttiker formula. These studies have further been supplemented by, e.g., exact transmission functions\cite{PhysRevB.83.064303}, heat transport models for refrigeration\cite{PhysRevB.86.125424}, nonlinear interactions\cite{PhysRevB.74.033408}, and thermal conductance studies in atomic junctions\cite{PhysRevE.75.061128,Hopkins2009,PhysRevB.86.235304,Bachmann2012}, graphene ribbons\cite{PhysRevB.79.115401, Xie2011, Yeo2012} and carbon nanotubes\cite{PhysRevLett.96.255503,Mingo2005}.

The nonequilibrium Green's function approach offers a natural framework to study transient effects both in electronic\cite{Kwong1998, Semkat1999, Dahlen2007} and phononic systems\cite{PhysRevB.81.052302, PhysRevE.82.021116, PhysRevE.86.031132}. Also in the case of phonon transport, it would be extremely useful to have a time-dependent Landauer--B{\"u}ttiker formula for interpreting the transient oscillations and relaxation times in an intuitive fashion. In addition, as the time-propagation of the equations of motion for the phonon Green's function is computationally a rather heavy task\cite{Sakkinen1,Sakkinen2}, bringing down the computational cost when studying larger systems would be advantageous. In this article, we present a substantial advance in this direction; we consider harmonic nanoscale systems connected to heat baths in a wide-band-like approximation, and derive a closed expression for the time-dependent phonon density matrix. In the steady-state limit our result reduces to a form of Landauer--B{\"u}ttiker type discussed in, e.g., Refs.~\onlinecite{Dhar2006,Dhar2012}.

We will first introduce the model and the dynamical equations of motion in our phonon transport setup (Sec.~\ref{sec:setup}), and then we will discuss the embedding scheme with its approximations and limitations (Sec.~\ref{sec:cutoff}). The solution to the equations of motion, the main result of the present work, is derived in Sec.~\ref{sec:eom} with more detailed calculations presented in the appendix. As an illustration of the derived formula we study the transient heat currents in small atomic chains (Sec.~\ref{sec:results}). A summary and main conclusions are drawn in Sec.~\ref{sec:conclusion}.


\section{Theoretical background}\label{sec:theory}
\subsection{Transport setup and assumptions}\label{sec:setup}
We model heat transport in a nanomechanical device coupled to phononic baths at different temperatures. The description is for noninteracting phonons only. The Hamiltonian for this setup can be written in terms of momentum and displacement field operators ($\hbar=1$)
\be
\hat{H} = \sum_j \frac{\hat{p}_j^2}{2m_j} + \sum_{jk}\frac{1}{2}\hat{u}_j K_{jk} \hat{u}_k
\ee
with indices $j,k = 1,\ldots,N$ running over the basis of the studied system, $m_j$ being the mass of the $j$-th oscillator, and $K$ being the positive definite force constant matrix. The momentum and displacement operators obey the canonical commutation relations $[\hat{u}_j,\hat{p}_k]=\im\delta_{jk}$, $[\hat{u}_j,\hat{u}_k] = 0 = [\hat{p}_j,\hat{p}_k]$. We define mass-normalized operators as $\hat{u}_j^\prime = \sqrt{m_j}\hat{u}_j$ and $\hat{p}_j^\prime = \hat{p}_j/\sqrt{m_j}$ obeying the same commutation relations, and write further
\be\label{eq:trfoham}
\hat{H} = \sum_j \frac{1}{2}(\hat{p}_j^\prime)^2 + \sum_{jk} \frac{1}{2}\hat{u}_j^\prime K_{jk}^\prime\hat{u}_k^\prime = \frac{1}{2}\sum_{jk,\mu\nu}\hat{\phi}_j^\mu\varOmega_{jk}^{\mu\nu}\hat{\phi}_k^\nu ,
\ee
where also the force constant matrix was transformed as $K_{jk}^\prime = K_{jk}/\sqrt{m_j m_k}$. The second equality follows by introducing the composite operators $\hat{\phi}$ with components $\hat{\phi}_j^1 = \hat{u}_j^\prime$ and $\hat{\phi}_j^2 = \hat{p}_j^\prime$. The indices $\{\mu,\nu\} \in \{1,2\}$ run over the different components of the field operators $\hat{\phi}$. The matrix elements of the block matrix $\varOmega_{jk}$ are given by $\varOmega_{jk}^{11} = K_{jk}^\prime$, $\varOmega_{jk}^{22} = \delta_{jk}$ and zero otherwise. The motivation behind the $\hat{\phi}$-``spinor''-representation is that the equations of motion are first order differential equations, instead of second order\cite{Sakkinen1,Sakkinen2}. The canonical commutation relations are encoded in the field operators $\hat{\phi}$ as
\be\label{eq:comm}
\left[\hat{\phi}_j^\mu,\hat{\phi}_k^\nu\right] = \delta_{jk}\alpha^{\mu\nu} \quad \text{with} \quad \alpha = \begin{pmatrix}0 & \im \\ -\im & 0\end{pmatrix} .
\ee
We accordingly define the phononic Green's function for time arguments $z$ and $z'$ on the Keldysh contour\cite{SvLBook} $\gamma$ as
\be\label{eq:def-green}
D_{jk}^{\mu\nu}(z,z') = -\im\langle\mathcal{T}_\gamma[\hat{\phi}_j^\mu(z)\hat{\phi}_k^\nu(z')]\rangle ,
\ee
where $\mathcal{T}_\gamma$ is the contour-time ordering operator and  $\langle \cdot \rangle$ is an ensemble average\cite{Schueler2016,Sakkinen1,Sakkinen2}. The equations of motion for the Green's function can be expressed through the time evolution of the field operators $\hat{\phi}$, and they read as (matrices in the $2N\times 2N$ representation are from now on denoted with boldface symbols)
\beq
\im\partial_z \bD_{jk}(z,z') & = & \alpha\delta_{jk}\delta(z,z') + \sum_{q}\alpha\bOmega_{jq}\bD_{qk}(z,z') \label{eq:d-eom-z}, \nonumber \\ \\
-\im\partial_{z'} \bD_{jk}(z,z') & = & \alpha\delta_{jk}\delta(z,z') + \sum_{q}\bD_{jq}(z,z')\bOmega_{qk}\alpha \label{eq:d-eom-zprime} . \nonumber \\
\eeq
Here each term is a $2\times2$ matrix.

\begin{figure}[t]
\centering
\includegraphics[width=0.45\textwidth]{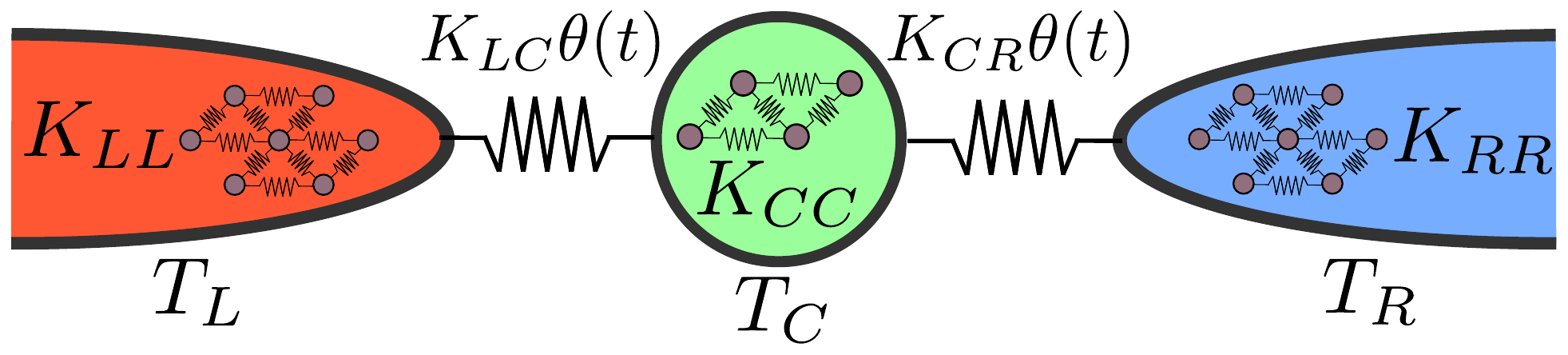}
\caption{(Color online) Heat transport setup where a central system of interest is connected to two reservoirs of different temperatures. The internal structures and couplings are defined by the force constant matrices $K$. The coupling refers to partitioned approach where the different systems are uncoupled at times $t<0$ and coupled at times $t\geq0$ when the system is driven out of equilibrium.}
\label{fig:setup}
\end{figure}

Let us look more specifically at the transport setup shown in Fig.~\ref{fig:setup}. The full Hamiltonian can be expressed as a composition of three parts: the left reservoir ($L$), the central system ($C$), and the right reservoir ($R$)
\be\label{eq:hammat}
\bOmega = \begin{pmatrix}
            \bOmega_{LL} & \bOmega_{LC} & 0 \\ 
            \bOmega_{CL} & \bOmega_{CC} & \bOmega_{CR} \\ 
            0 & \bOmega_{RC} & \bOmega_{RR} 
            \end{pmatrix} .
\ee
The different subsystems are coupled apart from a direct coupling between the reservoirs. In principle we have an underlying potential energy surface $E(\vec{R})$ for all nuclear coordinates $\vec{R}$ in the composite system which, in turn, determines the force constant matrix elements by $K_{jk} = \partial^2 E/\partial {R}_j \partial {R}_k$. For an uncoupled system we would only have the diagonal blocks nonzero in Eq.~\eqref{eq:hammat}. Here we consider only two reservoirs but the partitioning in Eq.~\eqref{eq:hammat} can directly be generalized to an arbitrary number of reservoirs to include, e.g., inner reservoirs as probes\cite{Dhar2006}. In Eq.~\eqref{eq:hammat} each diagonal block is a $2N_C \times 2N_C$ or $2N_\lambda \times 2N_\lambda$ matrix for $\lambda\in\{L,R\}$. The block structures for the diagonal elements are simply those discussed earlier
\beq
(\bOmega_{CC})_{j_C k_C} & = & \begin{pmatrix}K_{j_C k_C}^\prime & 0 \\ 0 & \delta_{j_C k_C}\end{pmatrix} , \\
(\bOmega_{\lambda\lambda})_{j_\lambda k_\lambda} & = & \begin{pmatrix}K_{j_\lambda k_\lambda}^\prime & 0 \\ 0 & \delta_{j_\lambda k_\lambda}\end{pmatrix} .
\eeq
The different regions, however, couple only through the displacement term, so the block structures for the off-diagonal elements are given by
\beq\label{eq:couplham}
(\bOmega_{C\lambda})_{j_C k_\lambda} & = & \begin{pmatrix}K_{j_C k_\lambda}^\prime & 0 \\ 0 & 0 \end{pmatrix} , \\ 
(\bOmega_{\lambda C})_{j_\lambda k_C} & = & \begin{pmatrix}K_{j_\lambda k_C}^\prime & 0 \\ 0 & 0 \end{pmatrix}
\eeq
for $\lambda\in\{L,R\}$.

We are mainly interested in the transport properties of the central system, so we extract the component corresponding to the central region, $CC$, from the equations of motion \eqref{eq:d-eom-z} and~\eqref{eq:d-eom-zprime}. This procedure leads to the following set of equations
\beq
& & (\bim_{CC}\partial_z - \balpha_{CC}\bOmega_{CC})\bD_{CC}(z,z') \nonumber \\
& = & \balpha_{CC}\delta(z,z') + \balpha_{CC} \bI_{CC}(z,z'), \qquad \label{eq:dprime-eom1}  \\
& &\bI_{CC}(z,z') = \int_\g \ud \bar{z} \bPi_{CC}(z,\bar{z})\bD_{CC}(\bar{z},z'), \label{eq:coll1}  \\
& & \bPi_{CC}(z,z') = \sum_{\lambda} \bOmega_{C \lambda}(z)\bd_{\lambda \lambda}(z,z')\bOmega_{\lambda C}(z'), \label{eq:emb-def} \\
& & (\bim_{\lambda\lambda}\partial_z - \balpha_{\lambda\lambda}\bOmega_{\lambda \lambda})\bd_{\lambda \lambda}(z,z') = \balpha_{\lambda\lambda} \delta(z,z'),\label{eq:small-d-eom} 
\eeq
and the corresponding adjoint equations. In the equations above $\balpha_{CC}\equiv \alpha \otimes \unity_{CC}$ and $\balpha_{\lambda\lambda} \equiv \alpha \otimes \unity_{\lambda\lambda}$ are $2N_C\times 2N_C$ and $2N_\lambda \times 2N_\lambda$ matrices, respectively. Here $\bd_{\lambda \lambda}$ is the isolated phonon Green's function in the reservoir $\lambda$ satisfying the equation of motion~\eqref{eq:small-d-eom} with the reservoir Hamiltonian $\bOmega_{\lambda\lambda}$. Since we are dealing with noninteracting phonons, the self-energy $\bPi_{CC}$ in the collision integral~\eqref{eq:coll1} is given solely as the embedding self-energy defined in terms of the coupling Hamiltonians $\bOmega_{C\lambda}$ in Eq.~\eqref{eq:emb-def}. From Eq.~\eqref{eq:dprime-eom1} and its adjoint we can derive an equation of motion for the \emph{lesser} Green's function $\bD_{CC}^<$ by using the Langreth rules for the collision integrals\cite{SvLBook}. Particularly, we are interested in the time-dependent one-particle reduced density matrix given by the time-diagonal $\brho(t) = \im\bD_{CC}^<(t,t)$. This is given by
\beq\label{eq:dlss-eom}
& & \im \frac{\ud}{\ud t}\bD^<(t,t) - \left[\balpha \bOmega \bD^<(t,t) - \bD^<(t,t)\bOmega \balpha\right] \nonumber \\
& = & -\{[\bD^{\text{R}} \cdot \bPi^<] + [\bD^< \cdot \bPi^{\text{A}}] + [\bD^\rceil \star \bPi^\lceil]\}(t,t)\balpha + \text{h.c.}, \nonumber \\
\eeq
where we defined $[a \cdot b](t,t) = \int_0^\infty \ud \bar{t} a(t,\bar{t})b(\bar{t},t)$ and $[a\star b](t,t) = -\im\int_0^\beta \ud \bar{\tau}a(t,\bar{\tau})b(\bar{\tau},t)$ for Keldysh functions $a$ and $b$, and we also dropped the subscripts $CC$ as we will only refer to the central region from now on\footnote{Strictly speaking the derivation of Eq.~\eqref{eq:dlss-eom} requires a single inverse temperature $\beta=1/T$ for the whole system because of the convolution along the imaginary axis. As we will consider partitioned systems, imaginary-time convolutions vanish and we can assign different temperatures to different subsystems.}. The Keldysh components lesser ($<$), greater ($>$), retarded ($\text{R}$), advanced ($\text{A}$), left ($\lceil$), and right ($\rceil$) of a function $k(z,z')$ on the contour are defined by\cite{SvLBook}
\beq
k^{\lessgtr}(t,t') & = & k(t_{\mp},t_{\pm}'), \nonumber\\
k^{\text{R}/\text{A}}(t,t')  & = & \pm\theta[\pm(t-t')][k^>(t,t')-k^<(t,t')], \nonumber\\
k^{\lceil}(\tau,t') & = & k(t_0-\im\tau,t'), \nonumber\\
k^{\rceil}(t,\tau) & = & k(t,t_0-\im\tau)\nonumber
\eeq
with the contour points $z=t_-$ on the forward branch, $z=t_+$ on the backward branch, and $z=t_0 - \im\tau$ on the vertical branch.

If we did not have the central system embedded into the environment, the self-energy terms in Eq.~\eqref{eq:dlss-eom} would simply be zero, and we would be left with a Liouville-type equation for the time-evolution of the reduced density matrix for an isolated central region. The self-energy terms, therefore, account for the open transport setup where a finite central region is embedded into the environment.

So far, the discussion has been rather general, and Eq.~\eqref{eq:dlss-eom} also applies to many different setups beyond the present study. For instance, interactions could be included by adding a many-body self-energy contribution. Our aim is to solve (analytically) this integro-differential equation for $\bD^<$ (in the equal-time limit) and then extract dynamical quantities such as heat currents from the time-dependent phonon  density matrix. Similar derivations have been done in earlier studies \cite{Stefanucci2004,Perfetto2008,Tuovinen2013,Tuovinen2014} for electron transport, and we will work along these previous ideas. For solving the equation we need to make some approximations, the first one being the \emph{partitioned} approach, i.e., all regions in the transport setup are initially ($t<0$) uncoupled and in separate thermodynamical equilibria. At $t=0$ we couple the different regions thus driving the system out of equilibrium, see Fig.~\ref{fig:setup}. From the perspective of an underlying potential energy surface, which we discussed earlier, this would lead to nonzero off-diagonal blocks in Eq.~\eqref{eq:hammat}. This coupling could, however, change the matrix elements in the diagonal blocks as well but we approximate them to be the same as for the uncoupled system.

The partitioning procedure disregards the initial couplings: $\bOmega_{\lambda C} = 0$ in equilibrium, so the integrations along the vertical track of the Keldysh contour on the right-hand side of Eq.~\eqref{eq:dlss-eom} are simply left out\cite{PhysRevB.81.052302,PhysRevE.82.021116}. The remaining two terms on the right-hand side of Eq.~\eqref{eq:dlss-eom} can be interpreted as a source/drain and a damping/equilibration term. The drain (source) term is a convolution between the propagator in the central region, $\bD^{\text{A}}$ (h.c.), and the lesser embedding self-energy $\bPi^<$ which is related to the probability of finding a phonon in the reservoirs, i.e., it describes the extraction (insertion) of phonons out of (into) the central region. The second term is a convolution between the propagator in the reservoirs, $\bPi^{\text{R}}$ (h.c.) and the lesser Green's function in the central region $\bD^<$ which is related to the probability of finding a phonon in the central region, i.e., it is responsible for damping (equilibration) effects.

For any type of reservoirs we may obtain $\bPi^{\text{R}}$ in a separate calculation. However, the complicated time-structure of the embedding self-energy makes it difficult to close the equations of motion into an analytically solvable form. Thus, we introduce an approximation. In the case of electronic transport, the commonly used wide-band approximation makes the embedding self-energy proportional to a delta function allowing for a closed solution\cite{Tuovinen2013,Tuovinen2014}. In the next section, we will make a similar approximation in the phononic case.

\subsection{Approximating the embedding self-energy}\label{sec:cutoff}
Let us start by considering the coupling of the central region to the reservoirs via the embedding self-energy. In order to evaluate the embedding self-energy from Eq.~\eqref{eq:emb-def} we need the coupling Hamiltonians and the reservoir Green's function, and, as we are considering the retarded component of the embedding self-energy, we need to find an expression for $\bd^{\text{R}}$. For the isolated phonon Green's function in the reservoir $\lambda$ we have the following expression
\beq\label{eq:d-terminal}
\bd_{\lambda\lambda}(z,z') & = & -\im\balpha_{\lambda\lambda}\theta(z,z')\bar{f}_\lambda(\bOmega_{\lambda\lambda}\balpha_{\lambda\lambda})\ex^{-\im\bOmega_{\lambda\lambda}\balpha_{\lambda\lambda}(z-z')} \nonumber \\
& - & \im\balpha_{\lambda\lambda}\theta(z',z)f_\lambda(\bOmega_{\lambda\lambda}\balpha_{\lambda\lambda})\ex^{-\im\bOmega_{\lambda\lambda}\balpha_{\lambda\lambda}(z-z')} \nonumber \\
\eeq
which can be checked to satisfy Eq.~\eqref{eq:small-d-eom} by calculating the derivative $\im\balpha_{\lambda\lambda}\partial_z \bd_{\lambda\lambda}(z,z')$. In the above expression $\bar{f}_\lambda = 1+f_\lambda$ and $f_\lambda(\w) = (\ex^{\beta_\lambda \w} - 1)^{-1}$ is the Bose--Einstein distribution for reservoir $\lambda$ at inverse temperature $\beta_\lambda = (k_{\text{B}}T_\lambda)^{-1}$. The expression in Eq.~\eqref{eq:d-terminal} is also proportional to the density matrix for an isolated system in the limit $z' \to z^+$. By using the above expression for the uncoupled Green's function we may derive different Keldysh components and calculate the retarded embedding self-energy from Eq.~\eqref{eq:emb-def}, see App.~\ref{app:secalc}. We find that the real and imaginary parts of the retarded embedding self-energy
\be\label{eq:pi-ret-mat}
\bPi_\lambda^{\text{R}}(\w) = \begin{pmatrix}\varPi_\lambda^{\text{R}}(\w) & 0 \\ 0 & 0 \end{pmatrix} = \bLambda_\lambda(\w) - \frac{\im}{2}\bGamma_\lambda(\w)
\ee
are, respectively, even and odd functions in frequency $\w$.

This finding also corresponds to the form calculated explicitly for a uniform one-dimensional system of $N$ coupled springs\cite{Wang2014}. We take this model for our reservoirs ($N_\lambda$ sites coupled with equal springs) and construct the embedding self-energy accordingly. In this model, the force constant matrix $K_{\lambda\lambda}^\prime$ has diagonal elements $2k_\lambda$ and the first off-diagonal elements $-k_\lambda$. In the limit $N_\lambda\to\infty$ the retarded embedding self-energy is given by $\varPi_\lambda^{\text{R}}(\w) = -k_\lambda z_\lambda(\w)$ where [see Sec.~7 in Ref.~\onlinecite{Wang2014}]
\beq\label{eq:z}
& & z_\lambda(\w) \nonumber \\
& = & \frac{1}{2k_\lambda}\left[2k_\lambda-\w^2+\zeta_\lambda(\w)\w\sqrt{(\w-2\sqrt{k_\lambda})(\w+2\sqrt{k_\lambda})}\right] \nonumber \\
\eeq
with $\zeta_\lambda(\w) = \text{sgn}(\w+2\sqrt{k_\lambda})$. Even though we have an explicit form for the frequency dependency of the embedding self-energy in Eq.~\eqref{eq:z}, we consider an approximation so that the equation of motion~\eqref{eq:dlss-eom} could be solved analytically. The full numerical solution of Eq.~\eqref{eq:dprime-eom1} is also possible using a time-stepping algorithm in the two-time plane\cite{Sakkinen1,Sakkinen2} but this is restricted to only small systems due to the computational cost.

We make \emph{a wide-band-like approximation} to Eq.~\eqref{eq:pi-ret-mat} so that for small frequencies (compared to other energy scales of the studied system)
\beq
\bLambda_\lambda(\w) & \approx & \bLambda_\lambda(\w=0) \equiv \bLambda_{0,\lambda}, \label{eq:linearized-re} \\ 
\bGamma_\lambda(\w) & \approx & \w \left(\partial_\w \bGamma_\lambda\right)_{\w=0} \equiv \w \bGamma_{0,\lambda}^\prime \label{eq:linearized-im} .
\eeq
In contrast to the conventional wide-band approximation in electronic transport, now the retarded embedding self-energy is not a purely imaginary constant. Instead, the real part is constant and the imaginary part is frequency dependent (linearized approximation). In fact, the imaginary part is not bounded when $|\w|\to\infty$ leading to some technical difficulties addressed soon. Also, since the force constant matrices $K$ are by construction positive definite, then from Eqs.~\eqref{eq:linearized-re} and~\eqref{eq:phonon-lambda} we see that $\bLambda_{0,\lambda}$ is negative definite, and from Eqs.~\eqref{eq:pi-ret-mat} and~\eqref{eq:linearized-im} we get that $\bGamma_{0,\lambda}^\prime$ is positive definite. Compared to Eqs.~\eqref{eq:linearized-re} and~\eqref{eq:linearized-im} similar wide-band-like approximations for the self-energy have been proposed in Refs.~\onlinecite{Galperin2004-1,Galperin2004-2} where the embedding self-energy is approximated as a purely imaginary sign function.

\begin{figure}[t]
  \centering
  \includegraphics[width=0.45\textwidth]{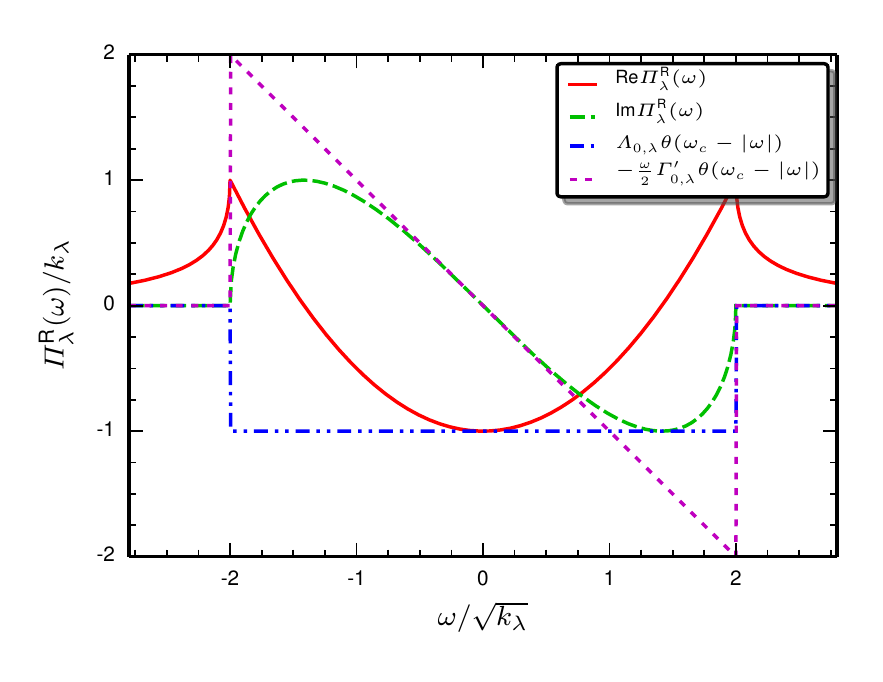}
  \caption{(Color online) Frequency dependency of the retarded embedding self-energy for the infinite coupled spring model: The solid red line is the real part and the long-dashed green line is the imaginary part. The respective approximations are the dash-dotted blue line and the short-dashed magenta line. The axes are scaled with the inter-atom force constant $k_\lambda$.}
\label{fig:phonon-sigma2}
\end{figure}

The real and imaginary parts of $\varPi_\lambda^{\text{R}}(\w)$ are plotted together with the approximations in Eqs.~\eqref{eq:linearized-re} and~\eqref{eq:linearized-im} [evaluated from Eq.~\eqref{eq:z}] in Fig.~\ref{fig:phonon-sigma2} versus $\w$. [In the figure we also employ a cut-off frequency, see Eq.~\eqref{eq:pi-retarded-w}.] 
Looking at the form in Eq.~\eqref{eq:z} the imaginary part of the self-energy is nonzero only for $|\w| < 2\sqrt{k_\lambda}$ introducing the phonon bandwidth. The motivation behind the approximations in Eqs.~\eqref{eq:linearized-re} and~\eqref{eq:linearized-im} around $\w=0$ is also clearly visible. 

Since the coupling matrices between different regions only have nonzero components in the $uu$ block, we express the embedding self-energy in the following form
\be\label{eq:linearized-sigma}
\bPi_\lambda^{\text{R}}(\w) \approx \begin{pmatrix} \varLambda_{0,\lambda}-\frac{\im\w}{2}\varGamma_{0,\lambda}^\prime & 0 \\ 0 & 0\end{pmatrix}  = \bLambda_{0,\lambda} - \frac{\im\omega}{2} \bGamma_{0,\lambda}^\prime ,
\ee
where each element of the $2\times 2$ matrix is an $N_C\times N_C$ matrix. Also, summing over the different reservoirs $\lambda$ gives the total embedding self-energy: $\bPi^{\text{R}} = \sum_{\lambda=L,R} \bPi_\lambda^{\text{R}}$, and this also applies to the real and imaginary parts $\bLambda_0$ and $\bGamma_0^\prime$. Compared to the conventional wide-band approximation in electronic transport, where the imaginary part of the self-energy is constant and the real part is zero, here we run into problems with our approximation for the embedding self-energy because of the unboundedness of the imaginary part $\w \bGamma_0^\prime / 2$ when $|\w|\to\infty$. This approximation for the real and imaginary parts of the embedding self-energy function does not satisfy the Kramers--Kronig relations\cite{Toll1956,SvLBook}, and for this reason we introduce a \emph{cut-off frequency} $\w_{\text{c},\lambda}$ above which the approximation for the embedding self-energy is simply zero
\be\label{eq:pi-retarded-w}
\bPi_\lambda^{\text{R}}(\w) = \theta(\w_{c,\lambda} - |\w|)(\bLambda_{0,\lambda} - \frac{\im\w}{2}\bGamma_{0,\lambda}^\prime) .
\ee
A natural choice for the cut-off frequency would be the phonon bandwidth, see Fig.~\ref{fig:phonon-sigma2}. We may tune the considered frequency range by varying the reservoir force constant so that the important processes are captured around comparatively small frequencies $\w/\sqrt{k_\lambda}$. The lesser component of the embedding self-energy is then simply given by the fluctuation--dissipation relation\cite{SvLBook}
\be\label{eq:pilss-step}
\bPi_\lambda^<(\w) = \theta(\w_{c,\lambda} - |\w|)[-\im f_\lambda(\w) \w \bGamma_{0,\lambda}^\prime] .
\ee
The cut-off frequency is a similar concept as the Debye temperature; here they are related by $\hbar\w_{c,\lambda} = k_{\text{B}}T_\lambda$\cite{Wang2007}. Instead of fixing the cut-off frequency at the phonon bandwidth we could also determine it by a frequency sum rule.

\subsection{Solving the equations of motion}\label{sec:eom}
Based on the approximation discussed in the previous section we derive expressions for the time convolutions in Eq.~\eqref{eq:dlss-eom}. More detailed calculations are shown in App.~\ref{app:conv} and we state here only the results. First, we define the effective (nonhermitian) Hamiltonian as
\be\label{eq:eff-ham}
\bOmega_{\text{eff}} = \frac{1}{\balpha+\frac{\im}{2}\bGamma_0^\prime}(\bOmega+\bLambda_0) 
\ee
which reduces to the uncoupled Hamiltonian, $\balpha\bOmega$, in the limit $\varLambda_0,\varGamma_0^\prime \to 0$. For cases relevant to us\footnote{E.g., in a one-site central region (scalar case) large values of $|\varLambda_0|$ compared to the vibrational frequency of the central region can break this analytical structure.}, this object has its eigenvalue spectrum in the lower-half plane, i.e., the imaginary part, given essentially by $\varGamma_0^\prime$, of the eigenvalues are negative. The real part of the eigenvalues occur at the eigenfrequencies of the uncoupled system shifted by $\varLambda_0$. For the adjoint $\bOmega_{\text{eff}}^\dagger$, on the other hand, the eigenvalues lie in the upper-half plane.

For the convolution between the retarded Green's function and lesser embedding self-energy we have
\beq\label{eq:phonon-1st-convolution}
& & \left[\bD^{\text{R}}\cdot \bPi^<\right](t,t) \nonumber \\
& = & -\im\sum_{\lambda=L,R}\int_{-\infty}^\infty \frac{\ud \w}{2\pi} \left[\unity-\ex^{\im(\w-\bOmega_{\text{eff}})t}\right]\bD^{\text{R}}(\w)\nonumber \\
& & \times\theta(\w_{c,\lambda} - |\w|) \w f_\lambda(\w) \bGamma_{0,\lambda}^\prime .
\eeq
As the frequency integral is cut off, we evaluate $\bD^{\text{R}}$ in the limit $\w_{c,\lambda}\to\infty$ [see Eq.~\eqref{eq:dr-specified}]:
\be\label{eq:ref-gr}
\bD^{\text{R}}(\w) = \frac{1}{\w-\bOmega_{\text{eff}}}\frac{1}{\balpha+\frac{\im}{2}\bGamma_0^\prime} .
\ee
By conjugating Eq.~\eqref{eq:phonon-1st-convolution}, we also get
\beq\label{eq:phonon-1st-convolution-dagger}
& & \left[\bPi^< \cdot \bD^{\text{A}}\right](t,t) = -\left[\bD^{\text{R}}\cdot \bPi^<\right]^\dagger(t,t) \nonumber \\
& = & -\im\sum_{\lambda=L,R}\int_{-\infty}^\infty \frac{\ud \w}{2\pi} \theta(\w_{c,\lambda} - |\w|) \w f_\lambda(\w) \bGamma_{0,\lambda}^\prime \nonumber \\
& & \times \bD^{\text{A}}(\w) \left[\unity-\ex^{-\im(\w-\bOmega_{\text{eff}}^\dagger)t}\right] ,
\eeq
where $\bD^{\text{A}}$ is found by conjugating Eq.~\eqref{eq:ref-gr}.

The convolution between the lesser Green's function and the retarded/advanced embedding self-energy is well-behaved for all cut-off frequencies. We therefore take the limit $\w_{c,\lambda} \to \infty$ in Eq.~\eqref{eq:pi-retarded-w}, see App.~\ref{app:conv} for more details, and obtain
\beq\label{eq:phonon-2nd-convolution}
\left[\bD^< \cdot \bPi^{\text{A}}\right](t,t) & = &  \bD^<(t,t)\bLambda_0 + \left.\partial_{t'} \bD^<(t,t')\right|_{t'=t}\frac{\bGamma_0^\prime}{2} ,\nonumber \\
\eeq
and by conjugating we also get
\beq\label{eq:phonon-2nd-convolution-dagger}
\left[\bPi^{\text{R}} \cdot \bD^<\right](t,t) & = & -\left[\bD^< \cdot \bPi^{\text{A}}\right]^\dagger(t,t) \nonumber \\
& = & \bLambda_0 \bD^<(t,t) + \frac{\bGamma_0^\prime}{2}\left.\partial_t \bD^<(t,t')\right|_{t=t'} . \nonumber \\
\eeq
The approximated form of the convolutions in Eqs.~\eqref{eq:phonon-1st-convolution}, \eqref{eq:phonon-1st-convolution-dagger}, \eqref{eq:phonon-2nd-convolution}, and~\eqref{eq:phonon-2nd-convolution-dagger} are expected to be in good agreement with a full Kadanoff--Baym propagation provided that the vibrational frequencies of the central region fall in the frequency window where $\Im\bPi^{\text{R}}(\w)$ is linear, see Fig.~\ref{fig:phonon-sigma2}. In Sec.~\ref{sec:validity} we assess the accuracy of the approximations by evaluating Eqs.~\eqref{eq:phonon-1st-convolution} and \eqref{eq:phonon-1st-convolution-dagger} with the exact $\bD^{\text{R}/\text{A}}$, and Eqs.~\eqref{eq:phonon-2nd-convolution} and \eqref{eq:phonon-2nd-convolution-dagger} with the finite cut-off $\bPi^{\text{R}/\text{A}}$.

An important thing to notice is that the the convolutions in Eqs.~\eqref{eq:phonon-2nd-convolution} and~\eqref{eq:phonon-2nd-convolution-dagger} depend not only directly on $\bD^<$ but also on the time-derivative of $\bD^<$ at equal-time limit. This means that inserting these expressions back into the equation of motion does not immediately close the equation for $\bD^<$. However, we may insert iteratively the explicit time-evolution from Eq.~\eqref{eq:dprime-eom1} and its adjoint for the derivative terms. As Eq.~\eqref{eq:dprime-eom1} includes a similar structure of time-convolutions, we will gain similar terms by the iteration procedure. Interestingly, it turns out that the iteration is truncated already after the first cycle due to the structure of the transport setup, and we get a closed equation for $\bD^<$.

Now we are ready to insert Eqs.~\eqref{eq:phonon-1st-convolution}, \eqref{eq:phonon-1st-convolution-dagger}, \eqref{eq:phonon-2nd-convolution}, and~\eqref{eq:phonon-2nd-convolution-dagger} into Eq.~\eqref{eq:dlss-eom}:
\begin{widetext}
\beq
& & \im \frac{\ud}{\ud t} \bD^<(t,t) - \left[\balpha\bOmega \bD^<(t,t)-\bD^<(t,t)\bOmega\balpha\right] \nonumber \\
& = & \balpha\left[-\im\sum_{\lambda=L,R}\intw\theta(\w_{c,\lambda} - |\w|)\w f_\lambda(\w)\bGamma_{0,\lambda}^\prime \bD^{\text{A}}(\w)\left[\unity-\ex^{-\im(\w-\bOmega_{\text{eff}}^\dagger)t}\right] + \bLambda_0 \bD^<(t,t) + \frac{\bGamma_0^\prime}{2}\left.\partial_t \bD^<(t,t')\right|_{t=t'} \right] \nonumber \\
& - & \left[\bD^<(t,t)\bLambda_0 + \left.\partial_{t'} \bD^<(t,t')\right|_{t'=t}\frac{\bGamma_0^\prime}{2} -\im\sum_{\lambda=L,R}\intw\theta(\w_{c,\lambda} - |\w|)\w f_\lambda(\w)\left[\unity-\ex^{\im(\w-\bOmega_{\text{eff}})t}\right]\bD^{\text{R}}(\w)\bGamma_{0,\lambda}^\prime\right]\balpha . \nonumber \\
\eeq
Then, we insert $\partial_t \bD(t,t')$ and $\partial_{t'} \bD(t,t')$ from Eq.~\eqref{eq:dprime-eom1} and its adjoint, and accordingly insert the consequent convolutions from Eqs.~\eqref{eq:phonon-1st-convolution}, \eqref{eq:phonon-1st-convolution-dagger}, \eqref{eq:phonon-2nd-convolution}, and~\eqref{eq:phonon-2nd-convolution-dagger}. This step generates a plethora of terms but we still expand all the parentheses to better see what is the overall structure of the various terms:
\beq
& & \im \frac{\ud \bD^<(t,t)}{\ud t} - \balpha\bOmega \bD^<(t,t) + \bD^<(t,t)\bOmega\balpha \nonumber \\
& = & -\im\balpha\sum_{\lambda=L,R} \intw\theta(\w_{c,\lambda} - |\w|) \w f_\lambda(\w){\bGamma}_\lambda^\prime \bD^{\text{A}}(\w)\left[1-\ex^{-\im(\w-\bOmega_{\text{eff}}^\dagger)t}\right] + \balpha {\bLambda}_0 \bD^<(t,t) - \im \balpha\frac{{\bGamma}_0^\prime}{2}\balpha\bOmega \bD^<(t,t) \nonumber \\
& - & \balpha\frac{{\bGamma}_0^\prime}{2}\balpha\sum_{\lambda=L,R} \intw\theta(\w_{c,\lambda} - |\w|) \w f_\lambda(\w){\bGamma}_\lambda^\prime \bD^{\text{A}}(\w)\left[1-\ex^{-\im(\w-\bOmega_{\text{eff}}^\dagger)t}\right] - \im\balpha\frac{{\bGamma}_0^\prime}{2}\balpha{\bLambda}_0 \bD^<(t,t) \nonumber \\
& - & \im\balpha\frac{{\bGamma}_0^\prime}{2}\balpha\frac{{\bGamma}_0^\prime}{2}\left.\partial_t \bD^<(t,t')\right|_{t=t'} - \bD^<(t,t){\bLambda}_0 \balpha - \im \bD^<(t,t)\bOmega\balpha \frac{{\bGamma}_0^\prime}{2}\balpha - \im \bD^<(t,t){\bLambda}_0 \balpha\frac{{\bGamma}_0^\prime}{2}\balpha \nonumber \\
& - & \im\left.\partial_{t'} \bD^<(t,t')\right|_{t'=t} \frac{{\bGamma}_0^\prime}{2}\balpha\frac{{\bGamma}_0^\prime}{2}\balpha -\sum_{\lambda=L,R}\intw\theta(\w_{c,\lambda} - |\w|) \w f_\lambda(\w)\left[1-\ex^{\im(\w-\bOmega_{\text{eff}})t}\right]\bD^{\text{R}}(\w){\bGamma}_\lambda^\prime \balpha\frac{{\bGamma}_0^\prime}{2}\balpha \nonumber \\
& + & \im \sum_{\lambda=L,R}\intw\theta(\w_{c,\lambda} - |\w|) \w f_\lambda(\w)\left[1-\ex^{\im(\w-\bOmega_{\text{eff}})t}\right]\bD^{\text{R}}(\w){\bGamma}_\lambda^\prime \balpha . \label{eq:all-terms}
\eeq
Next step would be to again insert the derivatives from Eq.~\eqref{eq:dprime-eom1} and its adjoint, but already in the above step of the iteration, all terms involving derivatives of the lesser Green's function vanish. We notice this truncation (and cancellation of other terms as well) by evaluating simple matrix products
\be
\bGamma_{0,(\lambda)}^\prime \balpha \bLambda_0 = \bLambda_0 \balpha \bGamma_{0,(\lambda)}^\prime = \bGamma_{0,(\lambda)}^\prime \balpha \bGamma_{0,(\lambda)}^\prime = 0 .
\ee
All higher order derivatives are also truncated based on these matrix structures. By combining accordingly and simplifying we end up with
\beq\label{eq:de}
& & \im \frac{\ud \bD^<(t,t)}{\ud t} - \bOmega_{\text{eff}}\bD^<(t,t) + \bD^<(t,t)\bOmega_{\text{eff}}^\dagger \nonumber \\
& = & -\im\sum_{\lambda=L,R} \intw \theta(\w_{c,\lambda} - |\w|) \w f_\lambda(\w)\left\{\balpha\bGamma_{0,\lambda}^\prime \bD^{\text{A}}(\w)\left[\unity-\ex^{-\im(\w-\bOmega_{\text{eff}}^\dagger)t}\right] - \left[\unity-\ex^{\im(\w-\bOmega_{\text{eff}})t}\right]\bD^{\text{R}}(\w)\bGamma_{0,\lambda}^\prime \balpha\right\}.
\eeq
\end{widetext}
In Eq.~\eqref{eq:de} we now have a linear, first order, nonhomogeneous differential equation for $\bD^<$ which can be solved uniquely with an initial condition. The solution is (see App.~\ref{app:sol})
\beq\label{eq:final-result}
\brho(t) & \equiv & \im \bD^<(t,t) \nonumber \\
& = & \im \bD_0^<(t,t)  \nonumber \\
& + & \sum_{\lambda=L,R} \int_{-\w_{c,\lambda}}^{\w_{c,\lambda}} \frac{\ud \w}{2\pi} f_\lambda(\w)\left[\unity-\ex^{\im(\w-\bOmega_{\text{eff}})t}\right] \nonumber \\
& & \qquad \times \ \bB_\lambda(\w)\left[\unity-\ex^{-\im(\w-\bOmega_{\text{eff}}^\dagger)t}\right] 
\eeq
with the initial condition
\be
\im \bD_0^<(t,t) = \ex^{-\im \bOmega_{\text{eff}}t}\balpha f_C (\bOmega\balpha)\ex^{\im \bOmega_{\text{eff}}^\dagger t}
\ee
stemming from the uncoupled lesser Green's function in the central region as in Eq.~\eqref{eq:small-dlss} where the distribution $f_C$ is defined via an equilibrium temperature for the central region before coupling. The spectral function $\bB_\lambda(\w) \equiv \bD^{\text{R}}(\w)\w\bGamma_{0,\lambda}^\prime \bD^{\text{A}}(\w)$ can be evaluated as
\beq\label{eq:spectral}
& & \bB_\lambda(\w) \nonumber \\
& = & \frac{1}{\w(\balpha+\frac{\im}{2}\bGamma_0^\prime)-\bOmega-\bLambda_0}\w\bGamma_{0,\lambda}^\prime \frac{1}{\w(\balpha-\frac{\im}{2}\bGamma_0^\prime)-\bOmega-\bLambda_0} \nonumber \\
& = & \frac{1}{\w-\bOmega_{\text{eff}}}(\bGamma_{0,\lambda}^\prime)_{\text{eff}}(\w)\frac{1}{\w-\bOmega_{\text{eff}}^\dagger}
\eeq
with $(\bGamma_{0,\lambda}^\prime)_{\text{eff}}(\w) = (\balpha+\im\bGamma_0^\prime/2)^{-1}\w\bGamma_{0,\lambda}^\prime(\balpha-\im\bGamma_0^\prime/2)^{-1}$. Eq.~\eqref{eq:final-result} is our main result for the time-dependent one-particle density matrix. Remarkably, it is a closed expression, i.e., no time propagation is needed for evaluating the time-dependent density matrix; this also is a general feature in previous studies including similar derivations\cite{Stefanucci2004,Perfetto2008,Tuovinen2013,Tuovinen2014,Ridley2015}. The transient behavior is encoded in the exponentials: We find oscillations $\w_{jk} = |\Re \w_{j,\text{eff}} - \Re \w_{k,\text{eff}}|$, where $\w_{\text{eff}}$ are the complex eigenvalues of the effective Hamiltonian $\bOmega_{\text{eff}}$, as transitions between the vibrational modes in the central region . Finally, $f_\lambda (\w)\bB_\lambda(\w)$ is well-behaving at $\w=0$ (although $f_\lambda(\w)$ diverges at zero), and the cut-off frequency $\w_{c,\lambda}$ regulates the nonintegrable behavior at $\w\to -\infty$.

It is instructive to investigate few limiting cases for Eq.~\eqref{eq:final-result}. At $t=0$ the square brackets vanish and we are left with the uncoupled result, as should be the case due to the initial condition. This also happens if the systems remain uncoupled, i.e., $\bLambda_0 = 0 = \bGamma_0^\prime$; then we are left with the free evolution of the initial state as $\bOmega_{\text{eff}} \to \balpha\bOmega$ and $\bB_\lambda(\w)\to 0$. The steady-state result comes from the limit $t\to\infty$ when the exponentials vanish due to the nonhermitian structure of $\bOmega_{\text{eff}}$ [see the discussion after Eq.~\eqref{eq:eff-ham}]: 
\beq\label{eq:ss-comp}
\brho_{\text{SS}} & = & \sum_{\lambda=L,R}\int_{-\w_{c,\lambda}}^{\w_{c,\lambda}} \frac{\ud \w}{2\pi}  f_\lambda(\w) \frac{1}{\balpha \w-\bOmega-(\bLambda_0-\frac{\im\w}{2}\bGamma_0^\prime)}\nonumber \\
& & \times\w\bGamma_{0,\lambda}^\prime \frac{1}{\balpha\w-\bOmega-(\bLambda_0+\frac{\im\w}{2}\bGamma_0^\prime)} .
\eeq
Within our self-energy approximation, $\bPi^{\text{R/A}}(\w) = \theta(\w_{c,\lambda}-|\w|)(\bLambda_0 \mp \frac{\im\w}{2}\bGamma_0^\prime)$, we may write Eq.~\eqref{eq:ss-comp} as
\be
\brho_{\text{SS}} = \sum_{\lambda=L,R}\int_{-\infty}^{\infty} \frac{\ud \w}{2\pi} f_\lambda(\w) \bD^{\text{R}}(\w)\bGamma_{\lambda}(\w) \bD^{\text{A}}(\w) .
\ee
Indeed, our time-dependent result reduces in the steady-state limit to similar ``Landauer'' type of results derived from various starting points\cite{Rego1998, Ozpineci2001, Haenggi2003, Dhar2006, Wang2006, PhysRevLett.96.255503, Dhar2008, Wang2008, Dhar2012}.

Finally, we point out that the integrals in Eq.~\eqref{eq:final-result} are possible to carry out analytically to some extent which considerably speeds up the computation. Similarly as in Ref.~\onlinecite{Tuovinen2014} this result can be expanded in the eigenbasis of the nonhermitian matrix $\bOmega_{\text{eff}}$. Furthermore, when the Bose function is expressed as a Pad{\'e} series\cite{Hu2010,Hu2011,RidleyPNGF6}, the resulting frequency integrals may be written in terms of complex logarithms and exponential integral functions.


\section{Results}\label{sec:results}
\subsection{Setup and quantities of interest}
Here we apply the derived result in Eq.~\eqref{eq:final-result} to study the transient behavior of the heat current in simple lattice models when coupled to reservoirs at different temperatures. We benchmark the validity of the approximation by comparing to full numerical solution to the equation of motion~\eqref{eq:dprime-eom1} with the embedding self-energy in Eqs.~\eqref{eq:pi-ret-mat} and~\eqref{eq:z}.

As the derived result provides information about the one-particle density matrix in the central region, we are interested in local quantities in this region. The local energy in the central region may be calculated as a sum over the $uu$ and $pp$ blocks of the product of the Hamiltonian and the density matrix
\be\label{eq:td-energy}
E(t) = \frac{\im}{2}\text{Tr}\left[\bOmega \bD^<(t,t)\right] .
\ee
The local heat current between the sites of the central region may be derived by considering the temporal change in local energy in a given site; this should amount to the sum of heat currents flowing in and out of that site\cite{PhysRev.132.168,PhysRevE.86.031107,PhysRevE.88.012128}. The local energy for site $j$ can be written as an expectation value $\eps_j = \langle \hat{H}_j\rangle$ of the local Hamiltonian $\hat{H}_j = [(\hat{p}_j')^2 + \sum_k\hat{u}_j' K_{jk}'\hat{u}_k']/2$. (This is chosen so that $\hat{H} = \sum_j \hat{H}_j$.) Then, for $\hat{H}$ being the total Hamiltonian for the central region [see Eq.~\eqref{eq:trfoham}], we get from the Heisenberg equation
\be\label{eq:local-energy}
\frac{\ud \eps_j}{\ud t}  = -\im \langle [\hat{H}_j , \hat{H}] \rangle = \sum_{k} \frac{1}{2}K_{jk}' \left(\langle \hat{u}_j'\hat{p}_k'\rangle - \langle \hat{u}_k'\hat{p}_j'\rangle\right),
\ee
where we used the commutation algebra of the momentum and displacement operators. This motivates to define the local (net) heat current between sites $j$ and $k$ as the $up$ component of the density matrix
\be\label{eq:heatcurrent}
J_{jk}^Q(t) = \frac{1}{2}K_{jk}^\prime \left(\langle \hat{u}_j' \hat{p}_k'\rangle - \langle \hat{u}_k' \hat{p}_j'\rangle\right)(t) ,
\ee
where the two terms can be regarded as ``in-coming'' (from $k$ to $j$) and ``out-going'' (from $j$ to $k$) heat current. This definition in Eq.~\eqref{eq:heatcurrent} deviates a little from the conventional definitions for the heat current between a reservoir and the central region\cite{PhysRevB.86.125424} since in our case there may be multiple (arbitrary) contacts between the sites contributing to the heat current in a given site.

We also note here that the above definition for a local energy (density) $\eps_j$ is ambiguous since we can add any local contribution $c_j$, for which $\sum_j c_j = 0$, which leaves the total energy unchanged. Further, the definition for the local heat current, obtained from the temporal change in local energy, is also not unique although widely used\cite{PhysRev.132.168,Prosen2000,Pereira2006,Arrachea2007,PhysRevE.86.031107,PhysRevE.88.012128}. The appropriate definition for the local energy density has recently been discussed in, e.g., Ref.~\onlinecite{Eich2014b}. Our motivation here, however, is to compare the derived analytical result in Eq.~\eqref{eq:final-result} with a full numerical solution, and this issue does not affect the comparison.

In our transport setup, we have uniform one-dimensional (semi-infinite) systems of coupled springs as reservoirs, and we fix the force constant in the reservoirs as $k_\lambda = 1$ for all leads and then relate the remaining parameters to this energy scale. Also, the central region, through which we study the heat current transients, is similarly a uniform one-dimensional (but finite) system of coupled springs for which we have the inter-atom force constant $k_C$:
\be\label{eq:kmatrix}
K' = \begin{pmatrix}
      2k_C & -k_C & \cdots & 0 \\
      -k_C & 2k_C & \cdots & \vdots \\
      \vdots & \vdots & \ddots & -k_C \\
      0 & \cdots & -k_C & 2k_C 
      \end{pmatrix}.
\ee
The terminal sites of the central region are coupled to the terminal sites of the reservoirs by the coupling force constant $k_{\lambda C}$.

As discussed in Sec.~\ref{sec:cutoff}, we can tune the embedding by choosing the coupling strength $k_{\lambda C}$. Also, for the vibrational frequencies of the central region to be inside the bandwidth (given by $\pm2\sqrt{k_\lambda}$) we choose the force constant in the central region $k_C$ small enough. As the force constants in the reservoirs are equal, the cut-off frequency also becomes $\lambda$-independent $\w_{c,L} = \w_{c,R} \equiv \w_c$. We also set the Boltzmann constant $k_{\text{B}} = 1$. We consider the temperature scale between the subsystems as a difference $\Delta T=T_L - T_R$ and set the temperature for the central region as $T_C = (T_L + T_R)/2$. We fix the temperature in the right reservoir $T_R=1$ and relate the remaining ones to this. After the systems are coupled the temperature in the central region loses its meaning during transient due to nonequilibrium conditions. Even if the system reaches a steady-state, it is not a thermal equilibrium but a nonequilibrium steady-state. Therefore, instead of ``thermalization'' we say that the system \emph{relaxes}, and we use a measure 
\be\label{eq:saturation}
\kappa = \frac{J^Q(t=\tau) - J^Q_{\text{SS}}}{J^Q_{\text{SS}}}
\ee
for defining \emph{the relaxation time} $\tau$ (time from $t=0$ to reach the steady state) as $\kappa = 10$\%\cite{Guo2015}. Evaluation of Eq.~\eqref{eq:saturation} also assumes the relative error to stay under the given tolerance at  times $t>\tau$.

As the structure of the central region and its couplings to the reservoirs are arbitrary, it is feasible to study also more complex systems, such as graphene ribbons\cite{PhysRevB.84.153412}, with the same methodology. Also, the nature of the heat transport (normal or anomalous\cite{Prosen2005,Lepri2016}) may be analyzed by different models with impurities or inner reservoirs. These type of simulations will, however, be postponed to future work since the focus of the present work is to benchmark the derived formula against exact results from the full propagation of the Kadanoff--Baym equations.


\subsection{Validity of the approximations}\label{sec:validity}
Compared to the full numerical solution of Eq.~\eqref{eq:dprime-eom1}, we have made a number of approximations when deriving the analytic solution in Eq.~\eqref{eq:final-result}. We study how much error each approximation causes when compared to the full numerical solution. First, we make the wide-band-like approximation for the self-energy in Eq.~\eqref{eq:pi-retarded-w} but otherwise we still solve the equation of motion numerically, i.e., without approximating the time-convolutions as was done in App.~\ref{app:conv} when deriving Eq.~\eqref{eq:final-result}. We denote this level of approximation as `WB'. Second, besides Eq.~\eqref{eq:pi-retarded-w}, we calculate the time-convolution $[\bD^{\text{R}}\cdot \bPi^<]$ as in Eq.~\eqref{eq:phonon-1st-convolution} using Eq.~\eqref{eq:ref-gr} (and similarly for the adjoint). This level of approximation is denoted as `WB-1'. Third approximation, we again use Eq.~\eqref{eq:pi-retarded-w} for the embedding self-energy but we calculate the time-convolution $[\bD^<\cdot \bPi^{\text{A}}]$ as in Eq.~\eqref{eq:phonon-2nd-convolution} using $\w_{c,\lambda}\to\infty$ in Eq.~\eqref{eq:pi-retarded-w} (and similarly for the adjoint). We refer to this approximation as `WB-2'. The main result in Eq.~\eqref{eq:final-result} follows when both WB-1 and WB-2 are made. Except for Eq.~\eqref{eq:final-result} the numerical solution of all the other schemes is obtained self-consistently in the full two-time plane\cite{Sakkinen1,Sakkinen2}.

We benchmark the validity of the approximations stated above by studying the heat current through a dimer molecule as the central region; the force constant matrix in Eq.~\eqref{eq:kmatrix} is then $2\times 2$. The parameters are chosen so that: (a) the vibrational frequencies of the dimer molecule are comparable with the cut-off frequency, i.e., the wide-band-like approximation is expected to fail; and (b) the vibrational frequencies of the dimer are in a narrow range compared to the reservoir bandwidth, i.e., the wide-band-like approximation should not neglect the detail of the spectrum. In case (a) we take $k_{\lambda C} = 1/2$ and $k_C = 1$, and in case (b) we take $k_{\lambda C} = 1/4$ and $k_C = 1/3$. In both cases we set the temperature profile by $T_L=5$, $\Delta T = 4$.

\begin{figure}[t]
\centering
\includegraphics[width=0.5\textwidth]{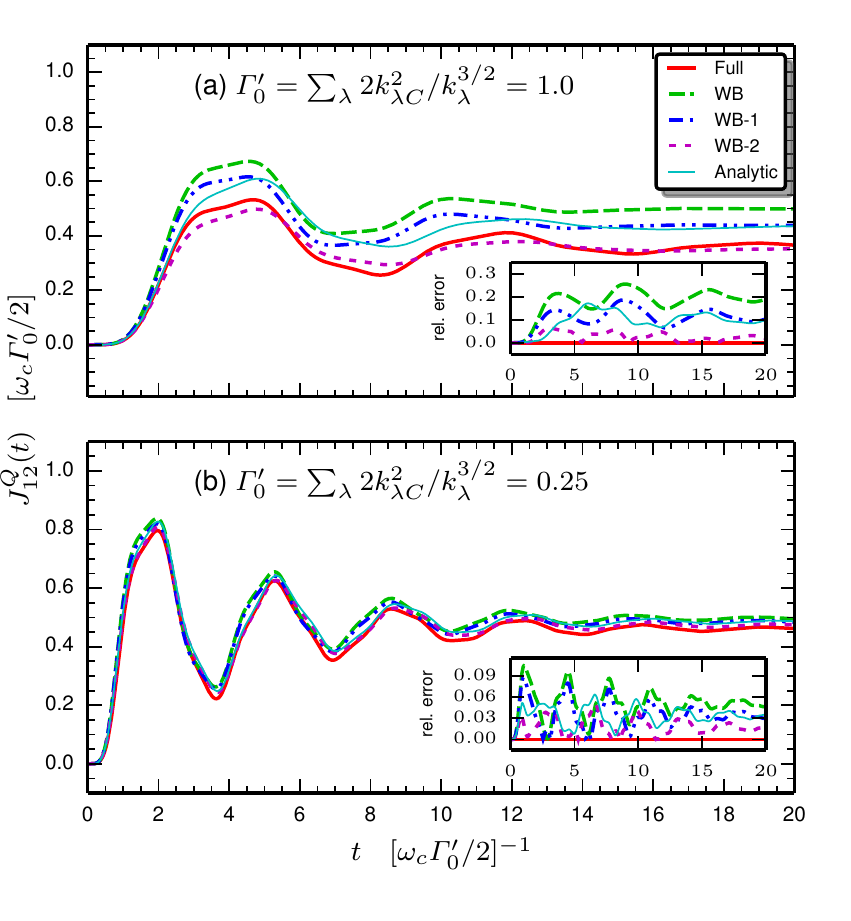}
\caption{(Color online) Time-dependent heat current through a dimer molecule (in units of appropriate energy scale) connected to two reservoirs with (a) strong coupling and broad spectrum, and (b) weak coupling and narrow spectrum. Thick solid red line is the numerical solution to Eq.~\eqref{eq:dprime-eom1}, long-dashed green, dash-dotted blue, short-dashed magenta, and thin solid cyan lines describe different levels of approximation, see text. The inset shows the difference between the approximate and full solutions.}
\label{fig:2site}
\end{figure}

\begin{figure}[t]
\centering
\includegraphics[width=0.5\textwidth]{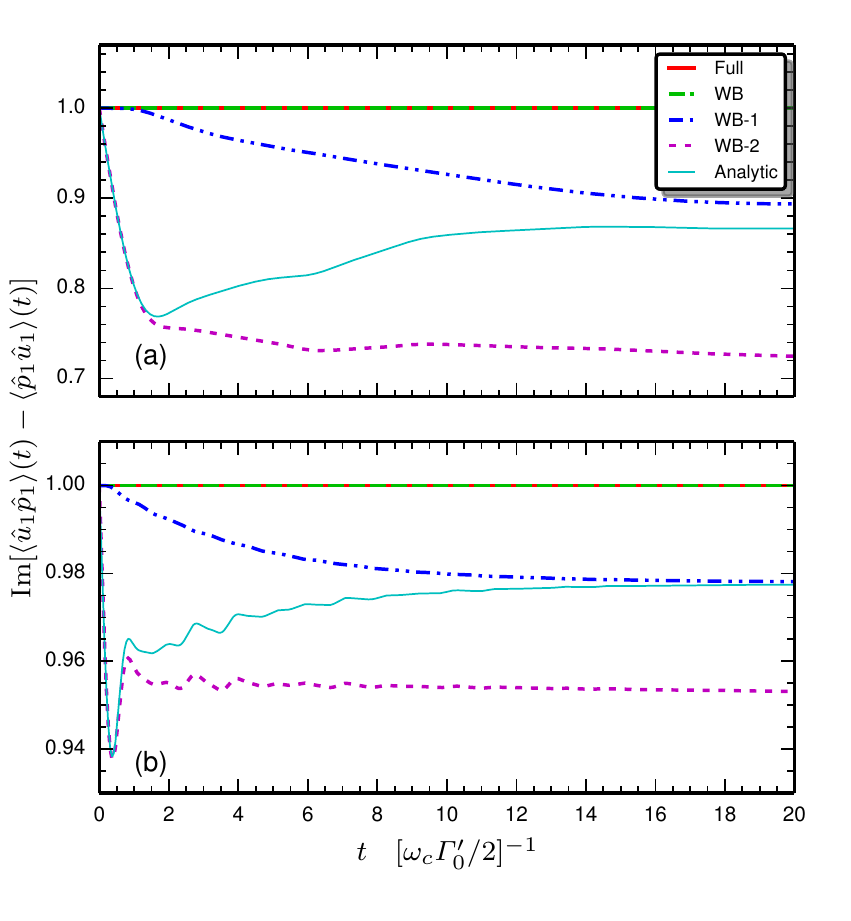}
\caption{(Color online) Time-dependence of the commutator between the displacement and momentum operators at site $1$ of a dimer molecule. The parameters of the calculation are the same as in Fig.~\ref{fig:2site}.}
\label{fig:commutator}
\end{figure}

In Fig.~\ref{fig:2site} we show the heat current, $J_{12}^Q$, between the dimer's atoms evaluated using Eq.~\eqref{eq:heatcurrent} for the two above-mentioned cases. In Fig.~\ref{fig:2site}(a) we see more deviation from the full solution (red, thick solid line) compared to Fig.~\ref{fig:2site}(b) as was expected from the parameter choice. By neglecting the frequency dependency of the phonon band in Eq.~\eqref{eq:z} (see Fig.~\ref{fig:phonon-sigma2}) we see in both panels (green, long-dashed curve, WB) that the transient behavior is overestimated due to too crude of an approximation for the band edges. (The approximated embedding self-energy abruptly drops to zero.) This also affects the long-time behavior as the current saturates towards a steady-state value faster than the full solution because the coupling strength (dissipation) is overestimated in WB. Also, the steady-state value for the heat current is overestimated. When we also add the approximation for the time-convolution $[\bD^{\text{R}}\cdot \bPi^<]$ (blue, dash-dotted curve, WB-1) we do not, at least in this case, see any qualitative difference to WB. This means that approximating the retarded Green's function in Eqs.~\eqref{eq:drdotpilss} and~\eqref{eq:dr-specified} as the embedded one only slightly modifies the initial transient and the steady-state values. When we consider the approximation only for the time-convolution $[\bD^<\cdot \bPi^{\text{A}}]$ (magenta, short-dashed curve, WB-2) we see a relatively good match with the full solution. However, individual density matrix elements may still differ considerably between the approximated and full solutions, see Fig.~\ref{fig:commutator} and Sec.~\ref{sec:timeshift}. When deriving $[\bD^{\text{R}}\cdot \bPi^<]$ in Eq.~\eqref{eq:phonon-1st-convolution-app} and $[\bD^<\cdot \bPi^{\text{A}}]$ in Eq.~\eqref{eq:phonon-2nd-convolution-app} we implicitly assumed the limit $\w_{c,\lambda}\to \infty$. Now, when the cut-off frequency is set to the phonon bandwidth, in Fig.~\ref{fig:2site}(a) the approximations do not fully take into account the broader spectrum of the central region. In Fig.~\ref{fig:2site}(b) the spectrum of the central region is more narrow, and all approximations give a reasonable agreement. Even if the general trend of the transient is qualitatively captured in both cases, quantitative differences can still be considerable, see the insets in Fig.~\ref{fig:2site}. When combining all the approximations, we get the analytic result in Eq.~\eqref{eq:final-result} (cyan, thin solid line). It is worth pointing out that this result, without needing any numerical evaluation of the Green's function, can give a comparatively good description.

To elucidate the transient behavior further we consider also the time-dependence of the commutator of the displacement and momentum operators $\langle[\hat{u}_1,\hat{p}_1]\rangle$ at site $1$; this is shown in Fig.~\ref{fig:commutator}. We see that this quantity in `Full' and `WB' equals the canonical commutation relation, $\im$, but when the time-convolutions are approximated in `WB-1' and `WB-2' we see a deviation. For these elements of the density matrix (from which the commutator is computed) this deviation is more significant than for the heat current in Fig.~\ref{fig:2site}. Further, this deviation can partly be tracked down to the approximation made for the retarded Green's function in Eq.~\eqref{eq:dr-specified}. From Eq.~\eqref{eq:def-green} and using $\bD^{\text{R}}(t,t') = \theta(t-t')[\bD^>(t,t') - \bD^<(t,t')]$ we should have
\be
\lim_{t\to t'^+} \bD^{\text{R}}(t,t') = -\im[\hat{\bphi},\hat{\bphi}] = -\im\balpha = \begin{pmatrix}0 & 1 \\ -1 & 0\end{pmatrix}
\ee 
with $\hat{\bphi}$ the composite field operators in Eq.~\eqref{eq:trfoham}, and the commutator is understood component-wise. (Each block in the above $2\times 2$ matrix is an $N_C \times N_C$ matrix.) However, if we directly Fourier transform Eq.~\eqref{eq:dr-specified} and take the equal-time limit, we end up with
\be
\lim_{t\to t'^+} \bD^{\text{R}}(t,t') = \begin{pmatrix}0 & 1 \\ -1 & -\varGamma_0^\prime/2\end{pmatrix} .
\ee
This means that the commutation relation, in case of the retarded Green's function, would only be satisfied in the limit of weak coupling, $\varGamma_0^\prime \to 0$. Fig.~\ref{fig:commutator}(b) shows that this is related to the commutation relation obtained from the lesser Green's function (density matrix elements), i.e., we observe a deviation of only few percent when the coupling is weak.

\subsection{Dependency on the cut-off frequency}\label{sec:cutoff-results}
\begin{figure}[t]
\centering
\includegraphics[width=0.5\textwidth]{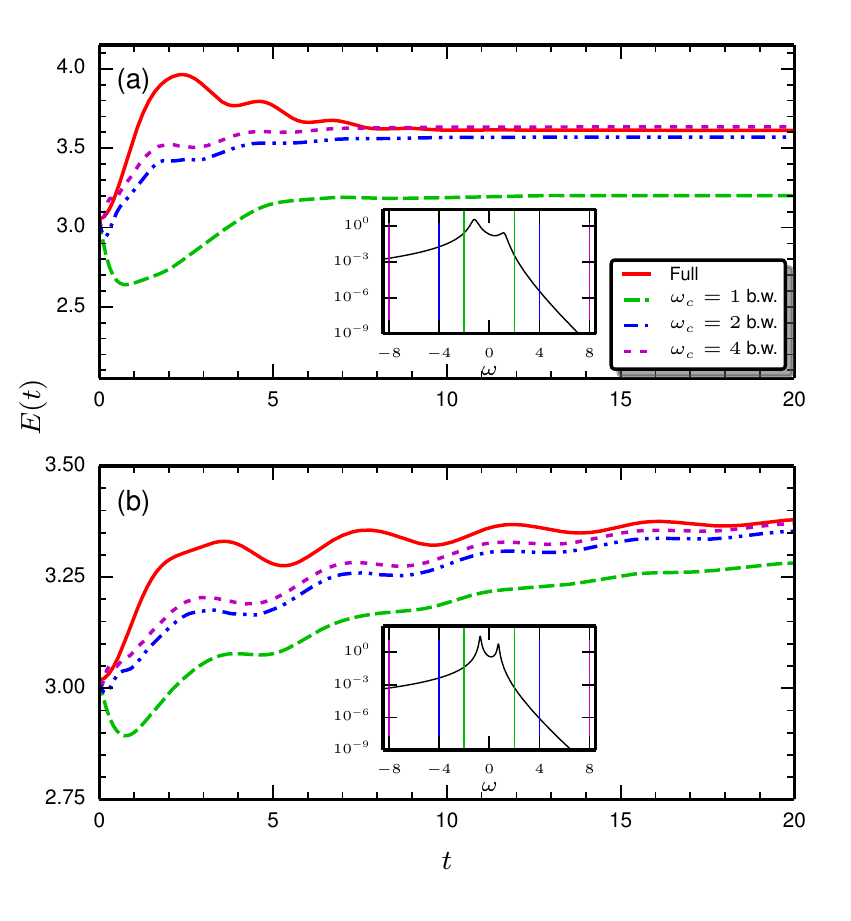}
\caption{(Color online) Time-dependent energy in a single site connected to two reservoirs with (a) strong coupling and broad spectrum, and (b) weak coupling and narrow spectrum. Solid red line is the numerical solution to Eq.~\eqref{eq:dprime-eom1}, dashed green, blue and magenta lines are obtained from Eq.~\eqref{eq:final-result} with different cut-off frequencies. The insets show the $uu$ component of $f_L(\w)\bB_L(\w)$ (in logarithmic scale) with vertical lines indicating the integration cut-off frequency.}
\label{fig:1site}
\end{figure}

We further compare the obtained analytic solution to the full numerical solution of the equation of motion~\eqref{eq:dprime-eom1}. We perform numerical integrations of Eq.~\eqref{eq:final-result} with varying cut-off frequency in terms of the bandwidth $2\sqrt{k_\lambda}$. In Fig.~\ref{fig:1site} we show the time-dependent energy in a single site connected to two reservoirs calculated from Eq.~\eqref{eq:td-energy}. We set $T_L=5$, $\Delta T = 4$, and study in Fig.~\ref{fig:1site}(a) strong coupling $k_{\lambda C} = 1/2$ with broad spectrum $k_C=1$, and in Fig.~\ref{fig:1site}(b) weak coupling $k_{\lambda C} = 1/4$ with narrow spectrum $k_C=1/3$. Both transients (the full solution) show similar features as the energy in the site first grows and oscillates and then saturates; reaching the steady-state takes longer in the weak coupling case, as is to be expected. In fact, since we are dealing with one site only, this quantity $E(t)$ can by the equipartition theorem be related to the time-dependent local temperature in the site\cite{Rubin1971,Lepri2003,PhysRevE.86.031107,PhysRevE.88.012128}. At $t=0$ the temperature is as prepared for the uncoupled system $T_C$, and after the transient the system saturates to a value within the temperature ``bias'' window $[T_L,T_R]$.

The insets in Fig.~\ref{fig:1site} show the frequency dependency of the $uu$ component of the steady-state integrand $f_L(\w)\bB_L(\w)$ in logarithmic scale and the integration limits corresponding to the cut-off values. As can be seen from the insets, the value of the integral depends on the cut-off frequency, and more specifically, on how well the spectral peaks in the central region fit in this frequency window. Also, as discussed earlier, the integral would not converge due to the self-energy approximation but it is regulated when $\w\to -\infty$ by the cut-off frequency. We see that in case (a) the integrand is broader and the spectral peaks are further away from $\w=0$ leading to a mismatch between the numerical solution with the full embedding self-energy and the derived time-dependent density matrix with the self-energy approximation. Increasing the cut-off frequency brings the time-dependent data closer to the full solution as less spectral weight is left outside of the integration region. It is still important to remember that if we do not limit the integration by the cut-off frequency the approximate solutions will grow (slowly) without a bound; in principle they do not approach the full solution. Due to the unsuitable parameters in case (a) the curves agree only close to the steady-state with higher cut-off values. In case (b) the match is better since the integrand is more narrowly peaked and the peaks are closer to $\w=0$. The oscillation frequencies correspond to the the peak separation in the spectral function. It can be seen that the transient frequency in the narrow spectrum case (b) is indeed smaller than in the broad spectrum case (a).

\begin{figure}[t]
\centering
\includegraphics[width=0.5\textwidth]{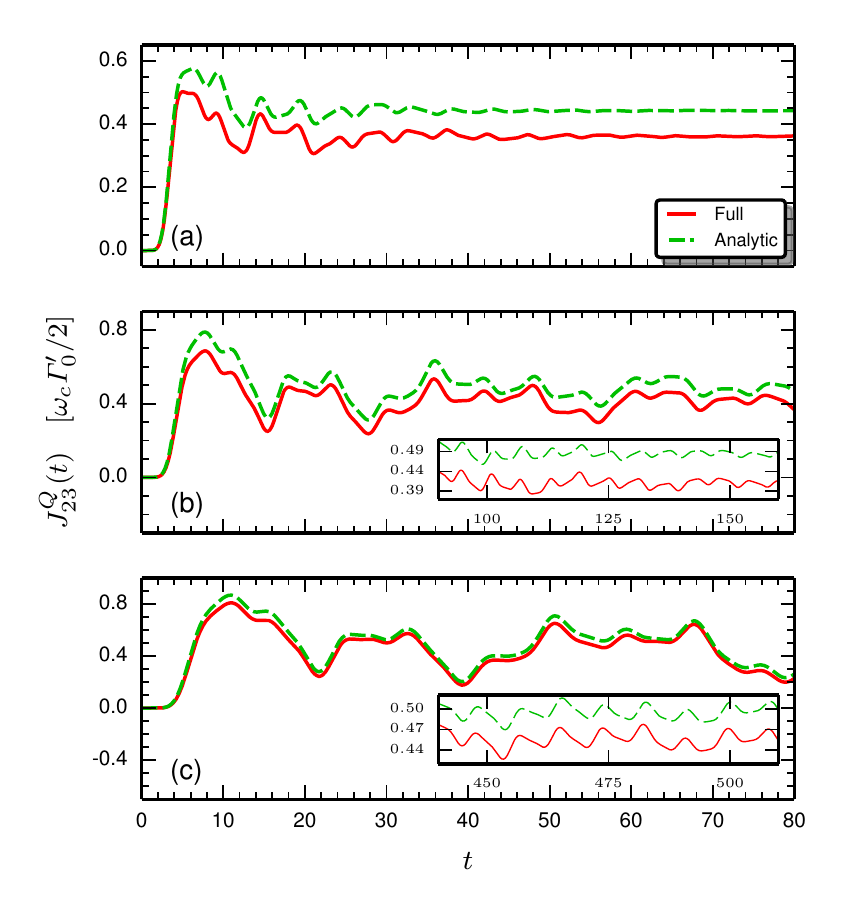}
\caption{(Color online) Time-dependent local heat current (in appropriate units) in the middle of a $4$-site chain connected to two reservoirs with (a) strong coupling and broad spectrum, (b) intermediate coupling and spectrum, and (c) weak coupling and narrow spectrum. Solid red line is the numerical solution to Eq.~\eqref{eq:dprime-eom1} and dashed green line is obtained from Eq.~\eqref{eq:final-result} by cut-off frequency at the phonon bandwidth, $\w_{c,\lambda}=2\sqrt{k_\lambda}$. The insets in panels (b) and (c) show the long-time behavior of the heat current.}
\label{fig:4site}
\end{figure}

\subsection{Heat current in a four-site system}
Next, we study the heat current in the middle of a $4$-site chain given by the local heat current in Eq.~\eqref{eq:heatcurrent} between sites $2$ and $3$, see Fig.~\ref{fig:4site}. Also here, we set the temperature difference as $T_L=5$, $\Delta T = 4$. In Fig.~\ref{fig:4site}(a) we have strong coupling $k_{\lambda C} = 1/2$ with broad spectrum $k_{C}=1$, in Fig.~\ref{fig:4site}(b) intermediate coupling and spectrum $k_{\lambda C} = 1/4$, $k_{C}=2/3$, and in Fig.~\ref{fig:4site}(c) weak coupling $k_{\lambda C} = 1/8$ and narrow spectrum $k_C=1/3$. Like in the previous example, we solve numerically the equation of motion~\eqref{eq:dprime-eom1} (solid red lines) and compare this to numerical integration of Eq.~\eqref{eq:final-result} (dashed green lines) by setting the cut-off frequency to the phonon bandwidth, $\w_{c,\lambda} = 2\sqrt{k_\lambda}$. 

Due to the strongest coupling in Fig.~\ref{fig:4site}(a) the steady-state is reached the fastest, but, as the spectrum is broad (reaching the edge of the band with tails outside the band) the transient is not captured very accurately. In Fig.~\ref{fig:4site}(b) the spectral peaks are inside the band although the broadening still causes differences with the full numerical result due to the cut-off frequency. Even though in Fig.~\ref{fig:4site}(b) most transient features are already captured, in Fig.~\ref{fig:4site}(c) the correspondence is even better since the spectrum is well inside the band with very narrow peaks. Also, in a longer time scale [see the insets in Figs.~\ref{fig:4site}(b) and~\ref{fig:4site}(c)], the transient oscillations can reach several hundred units of time before saturation. The approximate steady-state values overestimate the full solution by roughly $10$\%. More generally, we observe in steady-state the heat current to be positive, meaning a steady flow from the hot to the cold reservoir. However, if the temperature differences were smaller, the current could also momentarily (during the transient) flow from the cold to the hot reservoir (not shown). This finding is not related to the approximations (the effect was also seen in the full solution) but to the partitioned model with contacts being suddenly switched on\cite{PhysRevB.81.052302,PhysRevB.84.153412}; the effect would fade away in the case of adiabatic switching.

\subsection{Transient time shifts in individual density matrix elements}\label{sec:timeshift}
\begin{figure}[t]
\centering
\includegraphics[width=0.5\textwidth]{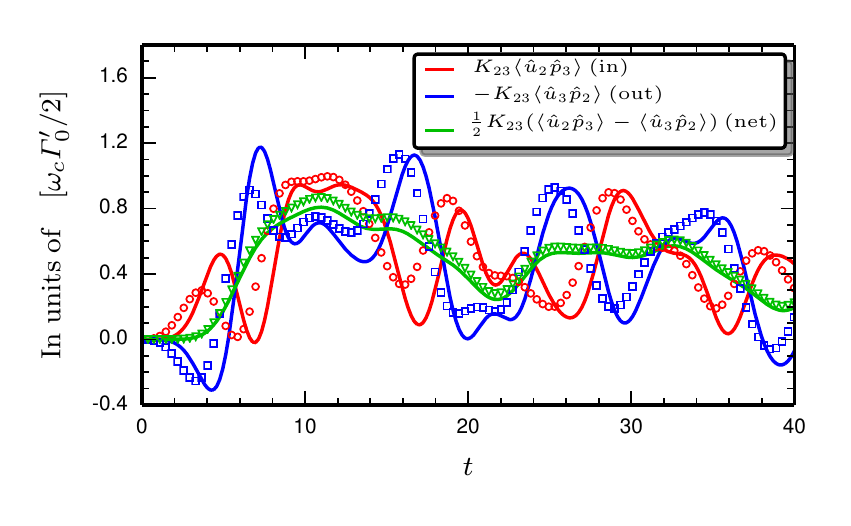}
\caption{(Color online) Time-dependence of the individual density matrix elements (real parts) when computing the heat current between sites $2$ and $3$ in a $4$-site chain, cf. Fig.~\ref{fig:4site}(c). The solid lines are the full numerical solutions and the markers with corresponding color are evaluated from Eq.~\eqref{eq:final-result}.}
\label{fig:timeshift}
\end{figure}

Even though the comparisons between the full numerical solution and the derived analytical result in Figs.~\ref{fig:2site} and~\ref{fig:4site} matched reasonably well, we may still wonder about the individual density matrix elements. This is due to the heat current in Eq.~\eqref{eq:heatcurrent} being an ``averaged-out'' quantity as it is evaluated as a sum of two contributions. In Fig.~\ref{fig:timeshift} we consider the same numerical simulation as in Fig.~\ref{fig:4site}(c) but now we plot separately the different terms in the definition of the heat current (and also the net current for reference). We notice that, even if the net current is described very accurately, the ``in-coming'' and ``out-going'' component of the current differ considerably (time shift) at short times between the full numerical solution (solid lines) and the analytic result in Eq.~\eqref{eq:final-result} (markers in corresponding color). This deviation relates to the approximation made for the time-convolution between the lesser Green's function and the advanced embedding self-energy; the short-time failure is considered more in detail in App.~\ref{app:conv}. If the cut-off frequency is not large enough compared to the relevant frequencies in the central region, the time shift becomes relatively more severe, see also Fig.~\ref{fig:commutator}.

This effect also relates to the fulfilment of the commutation relation investigated in Fig.~\ref{fig:commutator}. The approximations for the time-convolutions in deriving Eq.~\eqref{eq:final-result} result in  $up$ and $pu$ blocks of the density matrix possibly having complex elements when $j\neq k$ although they should be real-valued. This means that the heat current components, when considered separately as in Fig.~\ref{fig:timeshift}, gain a nonzero imaginary part. This imaginary part, however, cancels when evaluating the net heat current as a difference in Eq.~\eqref{eq:heatcurrent}.

If an accurate description is desired also for the individual matrix elements, it is possible as a ``rule of thumb'' to take the time shift into account by transforming the time axis for the approximate solution by $t\to t(1+a\ex^{-bt})$, where $a$ and $b$ are positive real numbers depending on the system parameters ($a=2\sqrt{\varGamma^\prime}$, $b=\w_{c,\lambda}\sqrt{\varGamma^\prime}/4$), to better represent the initial transient. We have tested (not shown) that during the initial transient the curves will be mostly in phase due to this shifting. Also, at larger times when the shifting goes to zero, the curves will stay in phase as argued in App.~\ref{app:conv}.

\subsection{Relaxation time scales}
\begin{figure}[t]
\centering
\includegraphics[width=0.5\textwidth]{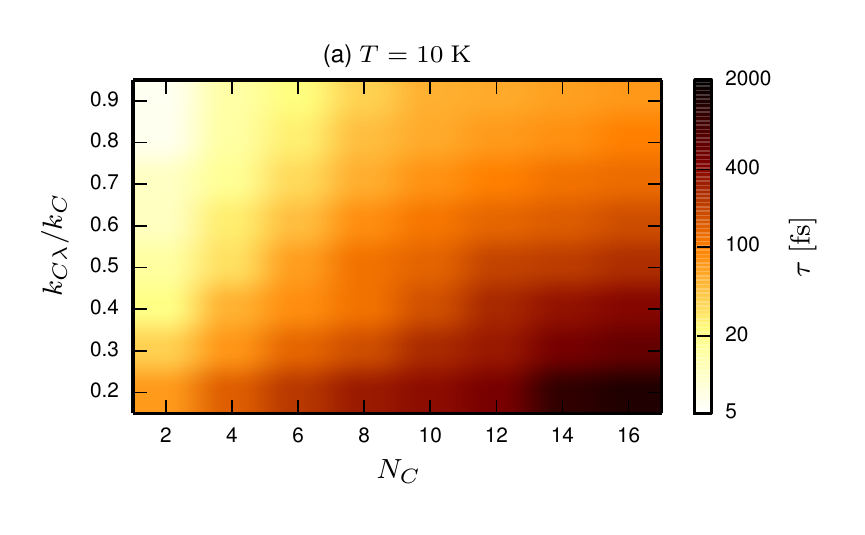}
~~~~~
\includegraphics[width=0.5\textwidth]{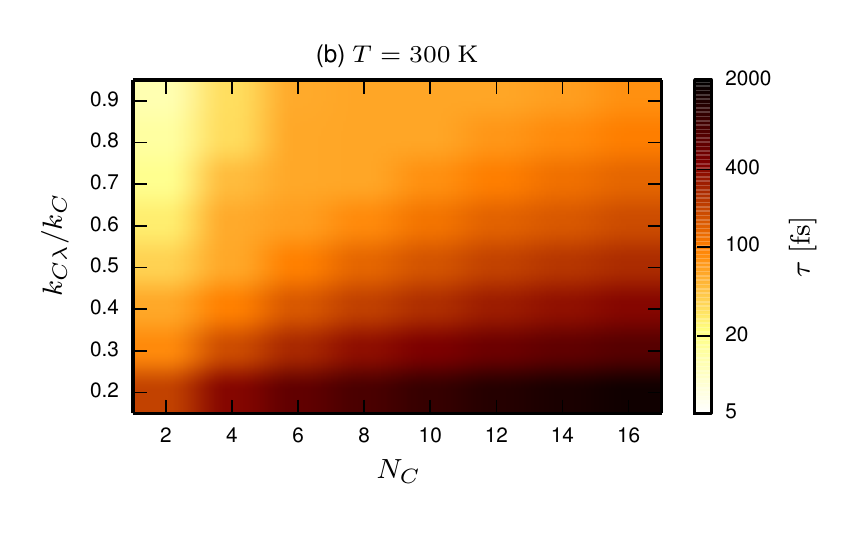}
\caption{(Color online) Relaxation times (colormap) for atomic chains of varying length (horizontal axis) and coupling strength to the reservoirs (vertical axis) at (a) low and (b) high temperature.}
\label{fig:saturation}
\end{figure}

We may also use the time-dependent formalism to estimate relaxation time scales using Eq.~\eqref{eq:saturation}. We consider similar atomic chains as in the previous sections but now we vary the length of the chain and the strength of the coupling between the chain and the reservoirs. We expect to see longer relaxation times for longer chains as it simply takes longer time for the wavefront to propagate over longer systems. Also, by increasing the coupling strength between the central region and the reservoirs the relaxation times should decrease due to the stronger dissipation. We choose the model parameters in physical units as $k_\lambda = 1{.}0$~eV/(\AA$^2$u), $k_C = 0{.}625$~eV/(\AA$^2$u) where u is the atomic mass unit. (Notice that these are the mass-normalized force constants, i.e., in SI units $[k]=1/\mathrm{s}^2$.) In Fig.~\ref{fig:saturation} we display the relaxation times (colormap) for chains of varying length (horizontal axis) and with varying coupling strength (vertical axis) having a $10$\% temperature difference between the reservoirs. In Fig.~\ref{fig:saturation}(a) the baseline temperature is $10$~K (leading to $T_L=11$~K and $T_R=9$~K), and in Fig.~\ref{fig:saturation}(b) we have the baseline temperature $300$~K (leading to $T_L=330$~K and $T_R=270$~K).

Indeed, we observe that, for both low and high temperature regimes, the relaxation is fastest in the shortest chain ($N_C=2$) when the coupling strength between the central region and the reservoirs is the highest ($k_{C\lambda} = 0.9k_C$). Expectedly, the slowest relaxation, which can take up to picosecond scale, occurs in the longest chain ($N_C=16$) when the coupling is the weakest ($k_{C\lambda} = 0.2k_C$). The overall dependency on the studied parameters is roughly similar in both low and high temperature regimes. However, we see in high temperature and small coupling that even the mid-range chains can take a comparatively long time for relaxation. On the other hand, increasing the coupling strength in the high-temperature regime leads to comparatively faster relaxation. This is also partly due to the larger absolute difference in temperature in the high-temperature case.

\section{Conclusion}\label{sec:conclusion}
We discussed a quantum transport model for noninteracting phonons in an arbitrary harmonic lattice setup. Using the nonequilibrium Green's function approach we derived an analytic result for the time-dependent one-particle reduced density matrix. The derivation involved a wide-band-like approximation for the embedding self-energy of the reservoirs; compared to the conventional wide-band approximation in electronic transport, now the real part of the embedding self-energy was set to be a nonzero constant around zero frequency whereas for the imaginary part we introduced a linear approximation. In addition, we introduced a cut-off frequency above which the embedding self-energy is simply set to zero. In the steady-state limit our analytical formula reduces to a known result\cite{Dhar2006,Dhar2012}.

As an application of the derived formula we analyzed transient heat currents in atomic chains. Using the numerical examples we were able to benchmark our derived result against a full numerical solution to the equations of motion for the Green's function, and furthermore test the validity of the approximations put forward when deriving the result. The approximations were found to be reasonable and the benchmark results congruent when the cut-off frequency was chosen large enough compared to the relevant energy scales of the studied systems, and when the coupling strength between the central region and the reservoirs was small enough for the wide-band-like approximation to hold.

Due to its computational simplicity, the introduced method holds promise for studying transient heat transport especially in large spatial scale, e.g., in graphene ribbon or carbon nanotube circuitries. It remains to be investigated if and how a partition-free approach, conventionally used in electronic transport\cite{Stefanucci2004}, could be incorporated also in the context of (time-dependent) phonon transport. A related issue, in the case of heat transport due to electrons, has recently been discussed in Refs.~\onlinecite{Eich2014a,Eich2016}, the latter concentrating on the transient regime.


\acknowledgments
We thank the Vilho, Yrj{\"o} and Kalle V{\"a}is{\"a}l{\"a} Foundation, the Academy of Finland, MIUR FIRB Grant No. RBFR12SW0, and EC funding through the RISE Co- ExAN (GA644076) for financial support. We further wish to acknowledge CSC -- IT Center for Science, Finland, for computational resources.


\appendix
\section{Self-energy calculations}\label{app:secalc}
From Eq.~\eqref{eq:d-terminal} we may deduce the greater and lesser components for the uncoupled Green's function to be
\beq
\bd_{\lambda\lambda}^>(t,t') & = & -\im\balpha_{\lambda\lambda}\bar{f}_\lambda(\bOmega_{\lambda\lambda}\balpha_{\lambda\lambda})\ex^{-\im\bOmega_{\lambda\lambda}\balpha_{\lambda\lambda}(t-t')} , \\
\bd_{\lambda\lambda}^<(t,t') & = & -\im\balpha_{\lambda\lambda} f_\lambda(\bOmega_{\lambda\lambda}\balpha_{\lambda\lambda})\ex^{-\im\bOmega_{\lambda\lambda}\balpha_{\lambda\lambda}(t-t')} , \label{eq:small-dlss}
\eeq
and further, the retarded component to read as
\beq
\bd_{\lambda\lambda}^{\text{R}}(t,t') & = & \theta(t-t')\left[\bd_{\lambda\lambda}^>(t,t') - \bd_{\lambda\lambda}^<(t,t')\right] \nonumber \\
& = & -\im\balpha_{\lambda\lambda}\theta(t-t')\ex^{-\im\bOmega_{\lambda\lambda}\balpha_{\lambda\lambda}(t-t')} .
\eeq
Since the retarded Green's function is a function of $t-t'$, we find by Fourier transforming
\be\label{eq:nonint-d}
\bd_{\lambda\lambda}^{\text{R}}(\w) = \balpha_{\lambda\lambda} \frac{1}{\w-\bOmega_{\lambda\lambda}\balpha_{\lambda\lambda}+\im\eta} = \frac{1}{\balpha_{\lambda\lambda}(\w+\im\eta)-\bOmega_{\lambda\lambda}} 
\ee
where, for the second equality, we used the idempotency of $\balpha$. The parameter $\eta$ is a positive infinitesimal accounting for the correct causal structure. We can then insert Eq.~\eqref{eq:nonint-d} into Eq.~\eqref{eq:emb-def} and derive an expression for the retarded self-energy for the central region embedded in the environment. We may evaluate the retarded Green's function by performing a matrix inversion for a block matrix resulting to
\begin{widetext}
\be\label{eq:reservoir-ret-green}
\bd_{\lambda\lambda}^{\text{R}}(\w) = \begin{pmatrix} \left((\w+\im\eta)^2 - K_{\lambda\lambda}^\prime\right)^{-1} & \im(\w+\im\eta)\left((\w+\im\eta)^2 - K_{\lambda\lambda}^\prime\right)^{-1} \\ -\im(\w+\im\eta)\left((\w+\im\eta)^2 - K_{\lambda\lambda}^\prime\right)^{-1} & K_{\lambda\lambda}^\prime\left((\w+\im\eta)^2 - K_{\lambda\lambda}^\prime\right)^{-1} \end{pmatrix}.
\ee
\end{widetext}
If we know the eigen decomposition of $K_{\lambda\lambda}^\prime$ as $K_{\lambda\lambda}^\prime X = X \w_\lambda^2$, we can then also diagonalize the full reservoir Hamiltonian with $\mathcal{X} \equiv \text{diag}(X, X)$ as
\be\label{eq:omega-onsite}
\widetilde{\bOmega}_{\lambda\lambda} = \mathcal{X}^\dagger \bOmega_{\lambda\lambda} \mathcal{X} = \begin{pmatrix}\w_\lambda^2 & 0 \\ 0 & 1 \end{pmatrix} \equiv \bw_\lambda^2 .
\ee
Further, we may write the retarded embedding self-energy in terms of the eigenmodes when inserting Eq.~\eqref{eq:reservoir-ret-green} into Eq.~\eqref{eq:emb-def}
\beq\label{eq:phonon-sigma-form}
(\bPi_\lambda^{\text{R}})_{j_C k_C}(\w)
& = & \sum_{q_\lambda}\widetilde{\bOmega}_{j_C q_\lambda}\frac{1}{(\w+\im\eta)^2-\w_{q_\lambda}^2}\nonumber\\
& & \quad \times \begin{pmatrix}1 & \im(\w+\im\eta) \\ -\im(\w+\im\eta) & \w_{q_\lambda}^2\end{pmatrix}\widetilde{\bOmega}_{q_\lambda k_C} \nonumber \\
\eeq
where we explicitly labelled the basis elements of the central region ($j_C , k_C$) and the reservoirs ($q_\lambda$). Also, $\widetilde{\bOmega}_{j_C q_\lambda} \equiv (\bOmega_{C \lambda}\mathcal{X})_{j_C q_\lambda}$.  Since the coupling Hamiltonians only have nonzero elements in the $uu$ block, this leads to the embedding self-energy having nonzero contribution also in the $uu$ block only, and we have
\beq\label{eq:pi-ret}
& & (\bPi_\lambda^{\text{R}})_{j_C k_C}^{11}(\w) \equiv (\varPi_\lambda^{\text{R}})_{j_C k_C}(\w) \nonumber \\
& = & \sum_{q_\lambda} \frac{\widetilde{K}_{j_C q_\lambda}\widetilde{K}_{q_\lambda k_C}}{(\w+\im\eta)^2-\w_{q_\lambda}^2} \nonumber \\
& = & \sum_{q_\lambda} \frac{\widetilde{K}_{j_C q_\lambda}\widetilde{K}_{q_\lambda k_C}}{2\w_{q_\lambda}}\left(\frac{1}{\w-\w_{q_\lambda}+\im\eta}-\frac{1}{\w+\w_{q_\lambda}+\im\eta}\right) , \nonumber \\
\eeq
where $\widetilde{K}_{j_C q_\lambda} \equiv (K_{C \lambda}^\prime{X})_{j_C q_\lambda}$. The advanced embedding self-energy $\varPi_\lambda^{\text{A}}$ is simply given by complex conjugating Eq.~\eqref{eq:pi-ret}. Then, we may evaluate the level-width function $\varGamma_\lambda$ defined as
\beq\label{eq:phonon-gamma}
& & (\varGamma_\lambda)_{j_C k_C}(\w) \equiv \im\left[(\varPi_\lambda^{\text{R}})_{j_C k_C}(\w) - (\varPi_\lambda^{\text{A}})_{j_C k_C}(\w)\right] \nonumber \\
& = & \im\sum_{q_\lambda}\frac{\widetilde{K}_{j_C q_\lambda}\widetilde{K}_{q_\lambda k_C}}{2\w_{q_\lambda}}\left(\frac{1}{\w-\w_{q_\lambda}+\im\eta}-\frac{1}{\w+\w_{q_\lambda}+\im\eta} \right.\nonumber \\
& & \left. \ -\frac{1}{\w-\w_{q_\lambda}-\im\eta}+\frac{1}{\w+\w_{q_\lambda}-\im\eta}\right) \nonumber \\
& = & \sum_{q_\lambda} \frac{\widetilde{K}_{j_C q_\lambda}\widetilde{K}_{q_\lambda k_C}}{\w_{q_\lambda}}\left(\frac{\eta}{(\w-\w_{q_\lambda})^2+\eta^2} \right.\nonumber \\
& & \left. \ -\frac{\eta}{(\w+\w_{q_\lambda})^2+\eta^2}\right) \nonumber \\
& = & \sum_{q_\lambda} \pi \widetilde{K}_{j_C q_\lambda}\frac{1}{\w_{q_\lambda}}\left[\delta(\w-\w_{q_\lambda})-\delta(\w+\w_{q_\lambda})\right]\widetilde{K}_{q_\lambda k_C} \nonumber \\
\eeq
where we used the lorentzian representation for the delta function $\pi\delta(x-a) = \lim_{\eta\to0}\eta/[(x-a)^2+\eta^2]$. Since $\varPi^{\text{A}} = (\varPi^{\text{R}})^\dagger$, we have that
\be
\varPi_\lambda^{\text{R}}(\w) = \varLambda_\lambda(\w) - \frac{\im}{2}\varGamma_\lambda(\w)
\ee
where $\varLambda_\lambda$ and $\varGamma_\lambda$ are real functions related by the Hilbert transform
\beq\label{eq:phonon-lambda}
& & (\varLambda_\lambda)_{j_C k_C}(\w) = \frac{1}{\pi}\mathcal{P}\int_{-\infty}^\infty\ud\w'\frac{(\varGamma_\lambda)_{j_C k_C}(\w')}{\w'-\w} \nonumber \\ 
& = & \sum_{q_\lambda} \frac{\widetilde{K}_{j_C q_\lambda}\widetilde{K}_{q_\lambda k_C}}{\w_{q_\lambda}}\left(\frac{1}{\w-\w_{q_\lambda}}-\frac{1}{\w+\w_{q_\lambda}}\right) .
\eeq
From Eqs.~\eqref{eq:phonon-gamma} and~\eqref{eq:phonon-lambda} we notice that $\varLambda_\lambda$ is an even function and $\varGamma_\lambda$ is an odd function.

\section{Time convolutions}\label{app:conv}
Here we calculate the convolutions $[\bPi^< \cdot \bD^{\text{A}}]$, $[\bD^{\text{R}} \cdot \bPi^<]$, $[\bPi^{\text{R}}\cdot \bD^<]$ and $[\bD^<\cdot \bPi^{\text{A}}]$ in Eq.~\eqref{eq:dlss-eom}.

We keep $\bD^{\text{R}}$ so far unspecified and we calculate the time convolution
\beq\label{eq:firstconv}
& & \left[\bD^{\text{R}}\cdot \bPi^<\right](t,t) = \int_0^\infty \ud \bar{t} \bD^{\text{R}}(t,\bar{t})\bPi^<(\bar{t},t) \nonumber \\
& = & \int_{-\infty}^\infty \frac{\ud \w}{2\pi} \int_{-\infty}^\infty \ud t' \bD^{\text{R}}(t') \ex^{\im\w t'} \bPi^<(\w) \theta(t-t') , 
\eeq
where we inserted the Fourier transform of $\bPi^<$, changed the integration variable as $t-\tb = t'$ and inserted a step function for extending the time interval to minus infinity. For the step function we may use the expression
\be
\theta(t-t') = \lim_{\eta \to 0^+} \int_{-\infty}^\infty \frac{\ud \w '}{2\pi \im} \frac{\ex^{\im\w ' (t-t')}}{\w' -\im \eta}
\ee
and evaluate further
\beq
& & \left[\bD^{\text{R}}\cdot \bPi^<\right](t,t)\nonumber \\
& = & \int_{-\infty}^\infty \frac{\ud \w}{2\pi}\int_{-\infty}^\infty\frac{\ud \bar{\w}}{2\pi\im}\frac{ \ex^{\im(\w-\bar{\w})t}\bD^{\text{R}}(\bar{\w})}{\w-\bar{\w}-\im\eta} \bPi^<(\w)  ,
\eeq
where we used the Fourier transform of $\bD^{\text{R}}(t)$ and changed the integration variable to $\w-\w' = \bar{\w}$. The exponential involving both frequencies can be split up, and we may also insert the approximation for the embedding self-energy
\beq\label{eq:drdotpilss}
& & \left[\bD^{\text{R}}\cdot \bPi^<\right](t,t) \nonumber \\
& = & \sum_{\lambda=L,R}\int_{-\infty}^\infty \frac{\ud \w}{2\pi} \ex^{\im\w t} \left[\int_{-\infty}^\infty \frac{\ud \w'}{2\pi \im} \frac{\ex^{-\im\w ' t}\bD^{\text{R}}(\w')}{\w-\w'-\im\eta}\right]\nonumber \\
& & \times \theta(\w_{c,\lambda} - |\w|)\left[-\im f_\lambda(\w) \w \bGamma_{0,\lambda}^\prime\right] .
\eeq
The cut-off frequency $\w_{c,\lambda}$ is now explicitly in this expression without specifying $\bD^{\text{R}}$. As we argued earlier, this expression is only valid in the cut-off regime and we simply use the embedded retarded Green's function for all frequencies $\w '$ (in the inner integral)
\be\label{eq:dr-specified}
\bD^{\text{R}}(\w) = \frac{1}{\balpha \w - \bOmega -\bPi^{\text{R}}(\w)} \approx \frac{1}{\w-\bOmega_{\text{eff}}}\frac{1}{\balpha+\frac{\im}{2}\bGamma_0^\prime} ,
\ee
where we inserted the approximation for the embedding self-energy and defined the effective Hamiltonian as in Eq.~\eqref{eq:eff-ham}. 

Now the retarded Green's function is specified, and $\bD^{\text{R}}$ and $\bPi^{\text{R}}$ satisfy the Dyson equation in the limits of $|\w| < \w_{c,\lambda}$. In Eq.~\eqref{eq:drdotpilss}, if the eigenvalues of $\bOmega_{\text{eff}}$ lie in the lower-half plane [see the discussion after Eq.~\eqref{eq:eff-ham}], then also the analytical structure for the Green's function is correct: $\bD^{\text{R}}(\w')$ has poles in the lower-half plane, and also the denominator goes to zero when $\w' = \w-\im\eta$ (in LHP). The key for evaluating the time-convolution was the approximation for $\bD^{\text{R}}$ in Eq.~\eqref{eq:drdotpilss}. As this is done within the cut-off frequency window $\theta(\w_{c,\lambda} -|\w|)$, we may analyze how large a difference does this approximation make compared to a full numerical solution of the equations of motion~\eqref{eq:dprime-eom1}. As we approximate the retarded Green's function for all frequencies $\w$ as in Eq.~\eqref{eq:dr-specified}, this means the limit $\w_{c,\lambda} \to \infty$. On the other hand, when we specify the cut-off frequency directly to the self-energy approximation in Eq.~\eqref{eq:pi-retarded-w}, this would amount to
\beq\label{eq:compare-dr}
\bD_{\theta}^{\text{R}}(\w) & \equiv & \frac{1}{\balpha \w -\bOmega - \bPi^{\text{R}}(\w)} \nonumber \\
& = & \frac{1}{\balpha \w -\bOmega - \theta(\w_{c,\lambda} - |\w|)(\bLambda_0 + \frac{\im\w}{2}\bGamma_0^\prime)} .
\eeq
As we discuss only the region $|\w| < \w_{c,\lambda}$ when evaluating the time-convolution in Eq.~\eqref{eq:drdotpilss}, we may compare how much the approximated Green's function deviates from that in Eq.~\eqref{eq:compare-dr} (outside the cut-off window)
\beq
\frac{\bD^{\text{R}}(\w)}{\bD_{\theta}^{\text{R}}(\w)} & = & \frac{\balpha \w -\bOmega - \theta(\w_{c,\lambda} - |\w|)(\bLambda_0 + \frac{\im\w}{2}\bGamma_0^\prime)}{\balpha \w - \bOmega - \bLambda_0 +\frac{\im\w}{2}\bGamma_0^\prime} \nonumber \\
& \stackrel{\w > \w_{c,\lambda}}{=} & \frac{\balpha \w - \bOmega}{\balpha \w -\bOmega - \bLambda_0 + \frac{\im\w}{2}\bGamma_0^\prime} \nonumber \\
& \xrightarrow{\w \gg \w_{\text{eff}}} & \frac{\balpha}{\balpha + \frac{\im}{2}\bGamma_0^\prime} \xrightarrow{\varGamma_0^\prime \ll 1} \unity
\eeq
where $\w_{\text{eff}}$ are the real parts of the eigenvalues of the effective Hamiltonian $\bOmega_{\text{eff}}$. This limit means: (1) we choose the cut-off frequency high enough so that the physical frequencies of the central region fall well inside this window; (2) if the frequency in the retarded Green's function still was higher than $\w_{c,\lambda}$, we would have, in the limit $\varGamma_0^\prime \to 0$ (weak coupling), that the relative difference in the retarded Green's functions approaches unity. Therefore, in order to make the approximation better is then two-fold: We can tune the force constant in the reservoirs so that the cut-off window is considerably broader than the energy scale of the central region; and/or we can also couple the central region more weakly to the reservoirs, thus, decreasing the value for $\varGamma_0^\prime$.

By using the eigenbasis of the effective Hamiltonian $\bOmega_{\text{eff}}$, and if the eigenvalues lie in the lower-half plane [again, also see the discussion after Eq.~\eqref{eq:eff-ham}], we can evaluate the integral in Eq.~\eqref{eq:drdotpilss} over $\w'$ by closing the contour in the lower-half plane ($t$ is a positive number):
\beq
& & \int_{-\infty}^\infty \frac{\ud \w '}{2\pi \im} \frac{\ex^{-\im\w ' t}}{\w-\w' -\im\eta} \frac{1}{\w' - \bOmega_{\text{eff}}}\frac{1}{\balpha + \frac{\im}{2}\bGamma_0^\prime} \nonumber \\
& = & -\left[\frac{\ex^{-\im\bOmega_{\text{eff}}t}}{\w - \bOmega_{\text{eff}}-\im\eta} - \frac{\ex^{-\im(\w-\im\eta)t}}{\w-\im\eta-\bOmega_{\text{eff}}}\right]\frac{1}{\balpha + \frac{\im}{2}\bGamma_0^\prime} . \nonumber \\
\eeq
Then we can take the limit $\eta\to 0$ and write 
\beq
\int_{-\infty}^\infty \frac{\ud \w '}{2\pi \im} \frac{\ex^{-\im\w ' t}\bD^{\text{R}}(\w')}{\w-\w' -\im\eta} & = & \frac{\ex^{-\im\w t} - \ex^{-\im\bOmega_{\text{eff}}t}}{\w-\bOmega_{\text{eff}}}\frac{1}{\balpha + \frac{\im}{2}\bGamma_0^\prime} \nonumber \\
& = & \left(\ex^{-\im\w t} - \ex^{-\im\bOmega_{\text{eff}}t}\right)\bD^{\text{R}}(\w) . \nonumber \\
\eeq
Inserting this into Eq.~\eqref{eq:drdotpilss} we get
\beq\label{eq:phonon-1st-convolution-app}
& & \left[\bD^{\text{R}}\cdot \bPi^<\right](t,t) \nonumber \\
& = & \sum_{\lambda=L,R}\int_{-\infty}^\infty \frac{\ud \w}{2\pi} \left[\unity-\ex^{\im(\w-\bOmega_{\text{eff}})t}\right]\bD^{\text{R}}(\w)\nonumber \\
& & \times \ \theta(\w_{c,\lambda} - |\w|)\left[-\im f_\lambda(\w) \w \bGamma_{0,\lambda}^\prime\right] . 
\eeq
It is worth noticing that this result could also be derived by Fourier transforming $\bD^{\text{R}}(\w)$ from Eq.~\eqref{eq:dr-specified} and then inserting the resulting $\bD^{\text{R}}(t,t')$ directly into the first row of Eq.~\eqref{eq:firstconv}. $[\bPi^< \cdot \bD^{\text{A}}]$ is found by conjugating Eq.~\eqref{eq:phonon-1st-convolution-app}.

Then we calculate the other time convolution in Eq.~\eqref{eq:dlss-eom}, $[\bD^< \cdot \bPi^{\text{A}}]$. From Eq.~\eqref{eq:pi-retarded-w} we get $\bPi^{\text{A}}(\w) = \left[\bPi^{\text{R}}(\w)\right]^\dagger$ and further in time domain
\beq
& & \bPi^{\text{A}}(t,t') = \intw \bPi^{\text{A}}(\w)\ex^{-\im\w(t-t')} \nonumber \\
& = & \intw \theta(\w_{c,\lambda} - |\w|)\left[\bLambda_0 + \frac{\im\w}{2}\bGamma_0^\prime\right]\ex^{-\im\w(t-t')} \nonumber \\
& = & \bLambda_0 \frac{1}{\pi(t-t')}\sin\left(\frac{t-t'}{1/\w_{c,\lambda}}\right) \nonumber \\
& - & \frac{\bGamma_0^\prime}{2}\partial_t\left[\frac{1}{\pi(t-t')}\sin\left(\frac{t-t'}{1/\w_{c,\lambda}}\right)\right] . \label{eq:sinc}
\eeq
In the limit $\w_{c,\lambda} \to\infty$ the sinc functions become delta functions, i.e., $\lim_{\epsilon\to 0} \frac{1}{\pi x}\sin\left(\frac{x}{\epsilon}\right) = \delta(x)$ and we obtain 
\be\label{eq:piadv-t}
\bPi^{\text{A}}(t,t') \to \bLambda_0\delta(t-t')-\frac{\bGamma_0^\prime}{2}\partial_t\delta(t-t') .
\ee
This is naturally the same result as if we put the limits of the integration in the derivation of Eq.~\eqref{eq:sinc} to $\pm \infty$. Based on the above expression for the advanced embedding self-energy we aim to calculate the time convolution in the equation of motion
\be\label{eq:dlss-piadv}
\left[\bD^< \cdot \bPi^{\text{A}}\right](t,t) = \int_0^\infty \ud \tb \bD^<(t,\tb)\bPi^{\text{A}}(\tb,t)
\ee
and the corresponding hermitian-conjugated one. The higher cut-off frequency $\w_{c,\lambda}$ we choose for the advanced embedding self-energy, the faster will the oscillations (in time) be in $\bPi^{\text{A}}(\tb,t)$. On the other hand, the fastest oscillations for $\bD^<(t,\tb)$ correspond to the (physical) vibrational frequencies of the central region. If we choose the cut-off frequency $\w_{c,\lambda}$ to be considerably higher than the typical energy scales in the central region, then in time-domain, $\bPi^{\text{A}}(\tb,t)$ appears \emph{almost as} Eq.~\eqref{eq:piadv-t} compared to $\bD^<(t,\tb)$. This allows us to calculate 
\beq\label{eq:phonon-2nd-convolution-app}
& & \left[\bD^< \cdot \bPi^{\text{A}}\right](t,t) \nonumber \\
& = & \int_0^\infty \ud \bar{t} \bD^<(t,\bar{t})\bPi^{\text{A}}(\bar{t},t) \nonumber \\
& = & \int_0^\infty\ud \bar{t} \bD^<(t,\bar{t})\sum_{\lambda}\left[\bLambda_{0,\lambda}\delta(\bar{t}-t)-\frac{\bGamma_{0,\lambda}^\prime}{2}\partial_{\bar{t}}\delta(\bar{t}-t)\right] \nonumber \\
& = &  \bD^<(t,t)\bLambda_0 + \left.\partial_{t'} \bD^<(t,t')\right|_{t'=t}\frac{\bGamma_0^\prime}{2} ,
\eeq
where we integrated by parts and noticed that the boundary term vanishes. By conjugating Eq.~\eqref{eq:phonon-2nd-convolution-app} we also find $\left[\bPi^{\text{R}} \cdot \bD^<\right]$. The result in Eq.~\eqref{eq:phonon-2nd-convolution-app}, however, implicitly assumes the limit $\w_{c,\lambda} \to \infty$ as we motivated its derivation by comparison of energy scales in $\bPi^{\text{A}}$ and $\bD^<$ exactly this way: $\w_{c,\lambda} \gg \w$ for frequencies $\w$ in the central region.

Let us try to justify this proposition by evaluating the time convolution of $\bPi^{\text{A}}$ and $\bD^<$ also by using the explicit expression in Eq.~\eqref{eq:sinc} and performing an asymptotic expansion in $\w_{c,\lambda}$ in
\beq\label{eq:splitup}
& & \left[\bD^< \cdot \bPi^{\text{A}}\right](t,t) = \int_0^\infty \ud \tb \bD^<(t,\tb) \bPi^{\text{A}}(\tb,t) \nonumber \\
& = & \int_0^t \ud \tb \bD^<(t,\tb)\bLambda_0\frac{1}{\pi(\tb-t)}\sin\left(\frac{\tb-t}{1/\w_{c,\lambda}}\right) \nonumber \\
& + & \int_0^t \ud \tb \bD^<(t,\tb) \frac{\bGamma^\prime}{2}\partial_{\tb}\left[\frac{1}{\pi(\tb-t)}\sin\left(\frac{\tb-t}{1/\w_{c,\lambda}}\right)\right] , \nonumber \\
\eeq
where the upper limit of the integration follows from the advanced nature of $\bPi^{\text{A}}(\tb,t)\sim\theta(t-\tb)$. Using Leibniz' rule we may write the second term of Eq.~\eqref{eq:splitup} as
\beq\label{eq:leibniz}
& & \int_0^t\ud\tb \bD^<(t,\tb)\frac{\bGamma^\prime}{2}\partial_{\tb}\left[\frac{1}{\pi(\tb-t)}\sin\left(\frac{\tb-t}{1/\w_{c,\lambda}}\right)\right] \nonumber \\
& = & \int_0^t\ud\tb\partial_t \bD^<(t,\tb)\frac{\bGamma^\prime}{2}\frac{1}{\pi(\tb-t)}\sin\left(\frac{\tb-t}{1/\w_{c,\lambda}}\right) \nonumber \\
& - & \frac{\ud}{\ud t}\left[\int_0^t \ud\tb \bD^<(t,\tb)\frac{\bGamma^\prime}{2}\frac{1}{\pi(\tb-t)}\sin\left(\frac{\tb-t}{1/\w_{c,\lambda}}\right)\right] .
\eeq
This tells us that we only need to consider ``the first line''-like terms in Eq.~\eqref{eq:splitup}
\be\label{eq:asymf}
F(t) = \int_0^t \ud \tb f(t,\tb) \frac{1}{\pi(\tb-t)}\sin\left(\frac{\tb-t}{1/\w_{c,\lambda}}\right) ,
\ee
and then we get ``the second line''-term in Eq.~\eqref{eq:splitup} by inserting $\partial_t \bD^<(t,\tb)\bGamma^\prime/2$ and $\bD^<(t,\tb)\bGamma^\prime/2$ as $f(t,\tb)$ and using Eq.~\eqref{eq:leibniz}. We then consider the behavior of the function $F$ in Eq.~\eqref{eq:asymf} when $\w_{c,\lambda}$ is large. We can get rid of the sinc function structure by taking the derivative with respect to $\w_{c,\lambda}$. (We assume we are allowed to differentiate under the integral sign since the functions are well-behaving in our case.) We obtain
\be
\frac{\ud F(t)}{\ud \w_{c,\lambda}} = \pi^{-1}\int_0^t \ud \tb f(t,\tb)\cos[\w_{c,\lambda}(\tb-t)] 
\ee
for which we may perform a sequential integration by parts since we know all the anti-derivatives of a cosine function
\beq
\frac{\ud F(t)}{\ud \w_{c,\lambda}} & = & \pi^{-1}\left\{f(t,\tb)\frac{1}{\w_{c,\lambda}}\sin[\w_{c,\lambda}(\tb-t)] \right.\nonumber \\
& + &\left. \partial_{\tb} f(t,\tb)\frac{1}{\w_{c,\lambda}^2}\cos[\w_{c,\lambda}(\tb-t)]\right.\nonumber \\
& - & \left.\partial_{\tb}^2 f(t,\tb)\frac{1}{\w_{c,\lambda}^3}\sin[\w_{c,\lambda}(\tb-t)]\right.\nonumber \\
& - & \left.\partial_{\tb}^3 f(t,\tb)\frac{1}{\w_{c,\lambda}^4}\cos[\w_{c,\lambda}(\tb-t)] + \ldots\right\}_0^t , \qquad
\eeq
where the remainder will only be higher order in $1/\w_{c,\lambda}$. This is justified by assuming that the derivatives $|\partial_{\bar{t}}f^{(k)}(t,\bar{t})|$ remain small compared to $\w_{c,\lambda}^{k+1}$ which is true as the oscillations in the function $f$ relate to the frequencies in the central region [see Eq.~\eqref{eq:splitup}] which are by construction assumed small compared to the cut-off frequency. Since we are interested in the large $\w_{c,\lambda}$ limit, we may simply take only the leading term, and integrating once over $\w_{c,\lambda}$ we obtain
\be\label{eq:asymres}
F(t) = \pi^{-1}\left\{f(t,\tb)\text{Si}[\w_{c,\lambda}(\tb-t)] + \mathcal{O}\left(\frac{1}{\w_{c,\lambda}^2}\right)\right\}_0^t
\ee
where $\text{Si}(x)$ is the \emph{sine integral}. Now we may consider Eq.~\eqref{eq:splitup} by inserting the corresponding parts as the function $f$ in Eq.~\eqref{eq:asymres} with the help of Eq.~\eqref{eq:leibniz}. Further, by using an asymptotic expansion for the sine integral\cite{NIST:DLMF} we may conclude that the terms neglected by approximating Eq.~\eqref{eq:splitup} by Eq.~\eqref{eq:phonon-2nd-convolution-app} are of the order $\mathcal{O}[(\w_{c,\lambda} t)^{-1}]$. Therefore, for long times the approximation is reasonable but even if we choose a large cut-off frequency $\w_{c,\lambda}$, for short times $t \lesssim 1/\w_{c,\lambda}$ the approximation fails.

\section{Integrating Eq.~\eqref{eq:de}}\label{app:sol}
It is convenient to start by making a transformation
\be
\bD^<(t,t) = \ex^{-\im \bOmega_{\text{eff}} t} \widetilde{\bD}^<(t,t)\ex^{\im \bOmega_{\text{eff}}^\dagger t} .
\ee
When evaluating the derivative of this product and canceling terms, the left-hand side of Eq.~\eqref{eq:de} simply becomes  $\ex^{-\im \bOmega_{\text{eff}} t} \im \frac{\ud\widetilde{\bD}^<(t,t)}{\ud t}\ex^{\im \bOmega_{\text{eff}}^\dagger t}$. We can then, accordingly, multiply both sides of the equation from left and right with the exponentials to get
\begin{widetext}
\beq
\frac{\ud \widetilde{\bD}^<(t,t)}{\ud t} & = & \sum_{\lambda=L,R}\intw \theta(\w_{c,\lambda}-|\w|)\w f_\lambda(\w) \ex^{\im \bOmega_{\text{eff}} t} \left[\bD^{\text{R}}(\w)\bGamma_{0,\lambda}^\prime\balpha - \balpha \bGamma_{0,\lambda}^\prime \bD^{\text{A}}(\w)\right] \ex^{-\im \bOmega_{\text{eff}}^\dagger t} \nonumber \\
& - & \sum_{\lambda=L,R} \intw \theta(\w_{c,\lambda}-|\w|) \w f_\lambda(\w)\left[\bD^{\text{R}}(\w)\bGamma_{0,\lambda}^\prime \balpha \ex^{\im(\w-\bOmega_{\text{eff}}^\dagger)t} - \ex^{-\im(\w-\bOmega_{\text{eff}})t}\balpha\bGamma_{0,\lambda}^\prime \bD^{\text{A}}(\w)\right] . \label{eq:diff-eq-d}
\eeq
\end{widetext}
Before we start integrating over $t$, recall the matrix structures 
\beq
\bD^{\text{R}}(\w) & = & \frac{1}{\w-\bOmega_{\text{eff}}}\frac{1}{\balpha+\frac{\im}{2}\bGamma_0^\prime}, \\
\bD^{\text{A}}(\w) & = & \frac{1}{\balpha-\frac{\im}{2}\bGamma_0^\prime}\frac{1}{\w-\bOmega_{\text{eff}}^\dagger}  .
\eeq
In Eq.~\eqref{eq:diff-eq-d} we have terms such as
\beq
\bD^{\text{R}}(\w)\bGamma_{0,\lambda}^\prime \balpha & = & \frac{1}{\w-\bOmega_{\text{eff}}}\frac{1}{\balpha+\frac{\im}{2}\bGamma_0^\prime} \bGamma_{0,\lambda}^\prime \balpha , \\
\balpha \bGamma_{0,\lambda}^\prime \bD^{\text{A}}(\w) & = & \balpha \bGamma_{0,\lambda}^\prime \frac{1}{\balpha-\frac{\im}{2}\bGamma_0^\prime}  \frac{1}{\w-\bOmega_{\text{eff}}^\dagger}
\eeq
where, conveniently, 
\be
\balpha \bGamma_{0,\lambda}^\prime \frac{1}{\balpha-\frac{\im}{2}\bGamma_0^\prime} = \frac{1}{\balpha+\frac{\im}{2}\bGamma_0^\prime} \bGamma_{0,\lambda}^\prime \balpha
\ee
which can be checked by simply evaluating the matrix products and inverses. Then, looking at the first row of Eq.~\eqref{eq:diff-eq-d}, we can perform the integration over $t$ by using the formula 
\beq
& & \int_0^t \ud t' \ex^{\im A t'}\left[\frac{1}{x-A}B-B\frac{1}{x-A^\dagger}\right]\ex^{-\im A^\dagger t'} \nonumber \\
& = & \left.-\im \ex^{\im A t'}\frac{1}{x-A}B\frac{1}{x-A^\dagger}\ex^{-\im A^\dagger t'}\right|_{0}^{t}
\eeq
for arbitrary matrices $A$ and $B$. The second row of Eq.~\eqref{eq:diff-eq-d} is simple to integrate over $t$ since there is only one exponential depending on time in each term. After integration we arrive at
\begin{widetext}
\beq\label{eq:solvediff}
\widetilde{\bD}^<(t,t) - \widetilde{\bD}^<(0,0^+) & = & 
-\im\sum_{\lambda=L,R} \intw \theta(\w_{c,\lambda}-|\w|) \w f_\lambda(\w)\left\{\ex^{\im\bOmega_{\text{eff}} t} \frac{1}{\w-\bOmega_{\text{eff}}}\frac{1}{\balpha+\frac{\im}{2}\bGamma_0^\prime}\bGamma_{0,\lambda}^\prime \frac{1}{\balpha-\frac{\im}{2}\bGamma_0^\prime}\frac{1}{\w-\bOmega_{\text{eff}}^\dagger} \ex^{-\im\bOmega_{\text{eff}}^\dagger t} \right.\nonumber\\
&  & \left. \hspace{20pt} + \frac{1}{\w-\bOmega_{\text{eff}}}\frac{1}{\balpha+\frac{\im}{2}\bGamma_0^\prime}\bGamma_{0,\lambda}^\prime \frac{1}{\balpha-\frac{\im}{2}\bGamma_0^\prime}\frac{1}{\w-\bOmega_{\text{eff}}^\dagger} \right. \nonumber \\
&  & \left. \hspace{20pt} - \frac{1}{\w-\bOmega_{\text{eff}}}\frac{1}{\balpha+\frac{\im}{2}\bGamma_0^\prime}\bGamma_{0,\lambda}^\prime \frac{1}{\balpha-\frac{\im}{2}\bGamma_0^\prime}\frac{1}{\w-\bOmega_{\text{eff}}^\dagger} \ex^{\im(\w-\bOmega_{\text{eff}}^\dagger)t} \right.\nonumber \\
&  & \left. \hspace{20pt} - \ex^{-\im(\w-\bOmega_{\text{eff}})t}\frac{1}{\w-\bOmega_{\text{eff}}}\frac{1}{\balpha+\frac{\im}{2}\bGamma_0^\prime}\bGamma_{0,\lambda}^\prime \frac{1}{\balpha-\frac{\im}{2}\bGamma_0^\prime}\frac{1}{\w-\bOmega_{\text{eff}}^\dagger} \right\}  .
\eeq
\end{widetext}
On the left-hand side we have the initial-state Green's function (at $t=0$). We are working in the partitioned scheme, i.e., the systems of different temperatures are coupled at $t=0$, so the initial condition should be equal to the uncoupled Green's function as in Eq.~\eqref{eq:small-dlss} but for the indices in the central region:
\be
\widetilde{\bD}^<(0,0^+) = \bD^<(0,0^+) = -\im\balpha f_C (\bOmega \balpha) .
\ee
Here $f_C$ gives the thermal distribution according to which the central system is prepared before it is connected to the reservoirs, and $\bOmega\balpha$ is correspondingly the uncoupled Hamiltonian. Now, in Eq.~\eqref{eq:solvediff} we may transform back from $\widetilde{\bD}^<(t,t)$ to $\bD^<(t,t)$ by multiplying with the exponentials from left and right and obtain our final result
\begin{widetext}
\beq
& & \im \bD^<(t,t) \nonumber \\
& = & \ex^{-\im \bOmega_{\text{eff}}t}\balpha f_C (\bOmega\balpha)\ex^{\im \bOmega_{\text{eff}}^\dagger t}  \nonumber \\
& + & \sum_{\lambda=L,R} \intw \theta(\w_{c,\lambda}-|\w|) \w f_\lambda(\w)\left[\unity-\ex^{\im(\w-\bOmega_{\text{eff}})t}\right] \frac{1}{\w-\bOmega_{\text{eff}}}\frac{1}{\balpha+\frac{\im}{2}\bGamma_0^\prime}\bGamma_{0,\lambda}^\prime \frac{1}{\balpha-\frac{\im}{2}\bGamma_0^\prime}\frac{1}{\w-\bOmega_{\text{eff}}^\dagger}\left[\unity-\ex^{-\im(\w-\bOmega_{\text{eff}}^\dagger)t}\right] \nonumber \\
\eeq
which can be written as Eq.~\eqref{eq:final-result} in main text by introducing the spectral function $\bB_\lambda(\w)$ in Eq.~\eqref{eq:spectral}.
\end{widetext}



%


\end{document}